\documentclass[AMA,STIX1COL]{WileyNJD-v2}

\usepackage{tablefootnote}
\usepackage{tabulary}
\usepackage{caption}
\usepackage{threeparttablex}
\usepackage{comment}
\usepackage{multirow}
\usepackage{arydshln}
\usepackage{colortbl}

\usepackage{lineno}

\usepackage{pdfpages}
\usepackage{pdflscape}
\usepackage{lscape}

\usepackage{hyperref}
\hypersetup{
    colorlinks=true,
    linkcolor=blue,
    filecolor=magenta,      
    urlcolor=cyan,
    pdftitle={Overleaf Example},
    pdfpagemode=FullScreen,
    citecolor=black,
    }
 
\articletype{Article Type}%
\received{xx xxxx xxxx}
\revised{xx xxxx xxxx}
\accepted{xx xxxx xxxx}
\begin{document}

\title{Application of targeted maximum likelihood estimation in public health and epidemiological studies: a systematic review}

\author[1]{Matthew J. Smith*}

\author[2]{Rachael V. Phillips}

\author[1,3]{Miguel Angel Luque-Fernandez$^{\pm}$}

\author[1]{Camille Maringe$^{\pm}$}

\authormark{Smith \textsc{et al.}}

\address[1]{\orgdiv{Inequalities in Cancer Outcomes Network}, \orgname{London School of Hygiene and Tropical Medicine}, \orgaddress{\state{London}, \country{England, United Kingdom}}}

\address[2]{\orgdiv{Division of Biostatistics, School of Public Health}, \orgname{University of California at Berkeley}, \orgaddress{\state{California}, \country{United States of America}}}

\address[3]{\orgdiv{Department of Statistics and Operations Research}, \orgname{University of Granada}, \orgaddress{\state{Granada}, \country{Spain}}}

\corres{*Matthew J. Smith. \newline
\email{matt.smith@lshtm.ac.uk}}

\presentaddress{London School of Hygiene and Tropical Medicine, Keppel Street, London, England, United Kingdom, WC1E 7HT \\
\newline 
$^{\pm}$ Both authors contributed equally to this work.}

\abstract[Abstract]{
\textbf{Background} The Targeted Maximum Likelihood Estimation (TMLE) statistical data analysis framework integrates machine learning, statistical theory, and statistical inference to provide a least biased, efficient and robust strategy for estimation and inference of a variety of statistical and causal parameters. We describe and evaluate the epidemiological applications that have benefited from recent methodological developments.
\newline
\textbf{Methods}
We conducted a systematic literature review in PubMed for articles that applied any form of TMLE in observational studies. We summarised the epidemiological discipline, geographical location, expertise of the authors, and TMLE methods over time. We used the Roadmap of Targeted Learning and Causal Inference to extract key methodological aspects of the publications. We showcase the contributions to the literature of these TMLE results.
\newline
\textbf{Results}
Of the 81 publications included, 25\% originated from the University of California at Berkeley, where the framework was first developed by Professor Mark van der Laan. By the first half of 2022, 70\% of the publications originated from outside the United States and explored up to 7 different epidemiological disciplines in 2021-22. Double-robustness, bias reduction and model misspecification were the main motivations that drew researchers towards the TMLE framework. Through time, a wide variety of methodological, tutorial and software-specific articles were cited, owing to the constant growth of methodological developments around TMLE.
\newline
\textbf{Conclusions}
There is a clear dissemination trend of the TMLE framework to various epidemiological disciplines and to increasing numbers of geographical areas. The availability of R packages, publication of tutorial papers, and involvement of methodological experts in applied publications have contributed to an exponential increase in the number of studies that understood the benefits, and adoption, of TMLE.
\newline
}


\maketitle


\section{Background}\label{sec1}

Public health decisions across many clinical specialities are often informed by research exploring the relationship between exposures and patient health outcomes. To ascertain a causal relationship, randomised control trials (RCT) are considered the gold standard because, through randomisation of subjects to a treatment, they reduce the possibility of bias. Observational data offer invaluable opportunities to study causal relationships in contexts where clinical trials might prove infeasible or unethical, as well as for studying groups of the population typically excluded from trials or beyond the initial target population. Under correct adjustment for selection bias, missingness, interference, and confounding, observational data complements the evidence coming from RCTs. \\

In both RCT and observational studies, a causal exposure-outcome relationship is presented with a causal estimand, such as the average treatment effect (ATE). Methodological statistical developments for causal inference attempt to produce the least biased estimate of the estimand and accurate inference. G-computation, Propensity Score (PS), and Inverse Probability of Treatment Weighting (IPTW) estimators rely on parametric modelling assumptions, which are susceptible to model misspecification (i.e., exclusion of the underlying and unknown data-generating distribution from the model, which can lead to biased estimates with small standard errors and thus misleading results). Double-robust methods, like Augmented Inverse Probability of Treatment Weighting (AIPTW) and Targeted Maximum Likelihood Estimation (TMLE), aim to minimise model misspecification by requiring estimation of both the outcome and exposure mechanisms. They provide a consistent estimator as long as either the outcome or exposure model is correctly specified. Double-robust methods often outperform single-robust methods in point and interval estimation.\cite{Smith2022,vanderLaan2011TargetedLearning} \\

TMLE, also known as targeted minimum loss-based estimation, was introduced by van~der~Laan and Rubin in 2006.\cite{vanderLaan2006} In general, TMLE is a two-step process that involves (1) initial estimation of the outcome and intervention models, and then (2) in a ``targeting'' step, uses information from them to optimise the bias-variance trade-off for the target estimand (e.g., ATE), rather than the whole outcome probability distribution. Furthermore, to avoid model misspecification, ensemble machine learning algorithms are used to estimate the initial models. In particular, the Super Learner (SL) algorithm for stacked ensemble machine learning is most commonly used as it is theoretically grounded and proven to perform optimally in large samples.\cite{VanDerLaan2007SuperLearner}\\
    
We lightly detail the technical steps involved in the TMLE of the ATE, i.e., the effect of a binary exposure $A$ on a post-exposure outcome $Y$, adjusted by baseline covariates $W$.\cite{Luque-Fernandez2018} The prediction function for the mean outcome $Y$, given exposure $A$ and covariates $W$ is estimated, most commonly, using SL. We could use this estimated prediction function, $\hat{E}[Y|A,W]$, to arrive at an estimate of the ATE. Specifically, we would obtain predicted outcomes under a counterfactual scenario where all subjects receive the exposure/treatment versus another scenario where no one receives it. The average difference between these predicted counterfactual outcomes is an estimate of the ATE. However, formal statistical inference (i.e., confidence intervals and p-values) cannot be obtained for this estimate and it is susceptible to residual confounding; the latter can be reduced by using the information on how each individual was assigned or allocated to each level of the exposure. We, therefore, estimate the function for predicting the probability of being exposed, given the covariates $W$, using SL (exposure model, i.e. propensity score). These first steps are common to other double-robust estimators of the ATE, such as AIPTW. We then calculate the so-called ``clever covariate'' for the ATE, which is the individual values of the binary exposure weighted by the predicted probabilities of the exposure, given $W$. This is similar to IPTW, except here we weight the predicted probability of each exposure level instead of the outcome. The fluctuation parameter ($\epsilon$) describes the difference between the observed outcome $Y$ and the initial predictions of the outcome from the outcome model. It is calculated through maximum likelihood estimation (MLE) by regressing the clever covariate on the observed outcome and incorporating as offset the predicted outcome values. When the fluctuation parameter is estimated as close to 0 there is little difference between the observed and predicted outcomes; thus, the propensity score does not provide additional information for the initial estimate of the outcome model because it was correctly specified. If the fluctuation parameter is not close to 0, then this indicates the presence of residual confounding in the initial estimate. The initial outcome model's predictions for each level of the binary exposure are updated using the fluctuation parameter $\epsilon$ as a weight, and the final ATE estimate is calculated from these updated estimates. The functional delta method based on the influence function can be used to derive the standard error of the ATE and construct Wald-type confidence intervals.  \\

Since 2006, the TMLE framework has experienced a growing number of theoretical and applied developments, and it has expanded further after a book that shared the TMLE framework to the international community of applied researchers was published in 2011.\cite{vanderLaan2011TargetedLearning} Targeting specifically applied researchers, efforts were made to provide lay-language descriptions of the framework and exemplify its applications.\cite{Gruber2015a,Schuler2017,Luque-Fernandez2018} Furthermore, in 2018, a second book was published disseminating more advanced applications of the TMLE framework to data scientists with a particular focus on longitudinal settings.\cite{vanderLaan2018TargetedLearning} TMLE is a robust framework for statistical analysis in clinical, observational and randomised studies. Since 2016, the American Causal Inference Conference has hosted a data challenge in which teams compete to estimate a causal effect in simulated data sets based on real-world data, such as from healthcare or education.\cite{UCBerkeley2022American2022} The competition is a proving ground for cutting-edge causal inference methods that have the potential to transform program evaluation. TMLE has consistently been a top-performing method.\cite{UCBerkeley2022AmericanResults} \\

The use of robust statistical methods is key to obtaining reliable results for public health and epidemiological research and maximising their benefit to society. Evidence shows that TMLE, by blending flexible machine learning methods and causal inference principles, is one such step towards robust causal claims that bear significant and practical effects. We reviewed the literature around public health and epidemiological applications of TMLE to date, alongside key TMLE developments over the last 20 years. We highlight the speed at which the field developed and spread through the scientific community, and identify areas for further development to increase the utility of the TMLE framework in epidemiological and applied research.

\section{Methods}\label{sec2}

\paragraph{Protocol registration and reporting standards} \,

    This study is reported using the Preferred Reporting Items for Systematic Reviews and Meta-Analyses (PRISMA) guideline. We registered this systematic review with PROSPERO (ID: CRD42022328482). 

\paragraph{Information sources} \,
    
    We searched the PubMed medical literature database for published epidemiological studies using TMLE in any epidemiological field (i.e., observational settings in biomedical sciences, including clinical epidemiology and public health). We searched for publications from any time up to 20th May 2022, the date the search was executed. The search strategy comprised of two specific groups of search terms focusing on TMLE and epidemiology. Relevant Mesh headings were included as were free-text terms, searched in the title, abstract or keyword fields. We used general and specific TMLE search terms, such as ``targeted maximum likelihood estimation'', ``targeted minimum loss-based estimation'', and ``targeted machine learning''. Epidemiological search terms included ``epidemiology'', ``public health'', ``population'', or ``treatment''. The two specific groups of terms were combined with ‘AND’ to retrieve the final set of results. Search strategies were developed with an information specialist (MALF). The full search strategy is shown in Table~\ref{tab:Table1}. 
    
    \begin{table}[ht]
    \centering
    \caption{Boolean search queries}
    \label{tab:Table1}
    \begin{tabular}{clr}
    \textbf{Query} & \multicolumn{1}{c}{\textbf{Boolean terms}} & \multicolumn{1}{c}{\textbf{Results}} \\ \hline
    \#1 & (epidemiology OR (public AND health) OR population OR treat*) & 11,004,506 \\
    \multicolumn{1}{l}{} &  & \multicolumn{1}{l}{} \\
    \#2 & \begin{tabular}[c]{@{}l@{}}("targeted maximum likelihood estimation") OR ("targeted minimum loss based estimation") OR \\ ("targeted minimum loss-based estimation") OR ("TMLE") OR ("targeted machine learning") OR \\ ("targeted learning") OR ("targeted machine-learning")\end{tabular} & 311 \\
    \multicolumn{1}{l}{} &  & \multicolumn{1}{l}{} \\
    \#3 & \#1 AND \#2 & 226 \\ \hline
    \end{tabular}
    \end{table}
    
\paragraph{Eligibility criteria} \,
    
    We excluded studies that did not report the use of TMLE as a tool to explore their (possibly causal) estimand of interest. We also excluded experimental studies, such as RCTs (n=15). We did not consider manuscripts that compared the performance of TMLE to other estimators, when there was no new development proposed and even if there was an applied question of interest, such as in Luque-Fernandez et al.\cite{Luque-Fernandez2018Data-AdaptivePresentation} Studies were restricted to the English language and primary research studies. Secondary research studies, such as reviews and comments of TMLE, conference abstracts and brief reports, and preprints were not searched. We classified the retained manuscripts into observational, methodological and tutorial articles. TMLE methodological development articles and tutorials were considered separately, even if they contained a methodological development specifically designed to investigate an epidemiological question within the same article. We make reference to these methodological articles throughout this review, as they underpin the applied publications. 
    
\paragraph{Study selection, data extraction and management} \,
    
\begin{figure}[ht]
        \centering
        \includegraphics[width=0.6\textwidth]{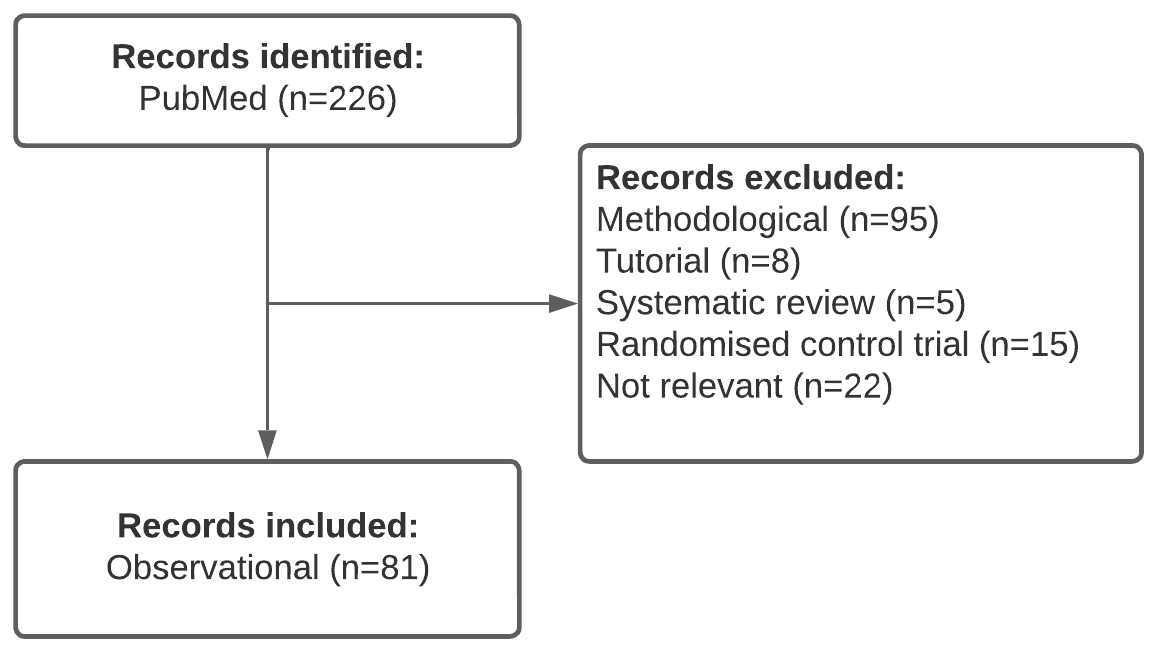}
        \caption{Flow diagram of studies included in the systematic review}
        \label{fig:flowchart}
    \end{figure}

    All retrieved publications were imported into the Endnote reference software where they were checked for duplication. Two of the three lead researchers (authors MJS, MALF and CM) were randomly allocated two-thirds of the 226 articles, to screen titles and abstracts of each publication independently and classify these into (1) observational, (2) methodological developments, (3) tutorial, (4) systematic review, (5) RCT, or (6) not relevant (Figure \ref{fig:flowchart}). Disagreements were discussed and adjudicated by the third independent reviewer, where necessary. Two researchers (authors MJS and CM) independently reviewed the full text of all eligible observational publications for data extraction. \\
    
\newpage

\section{Results}\label{sec3}

    We found 226 unique publications published prior to 20th May 2022 in PubMed (Figure \ref{fig:flowchart}). Of these, 95 articles were a methodological development (including theoretical - or software-based), eight were tutorials, five were systematic reviews, and fifteen were RCTs. Of the 22 articles that were not relevant, three mentioned ``TMLE'' only in the author fields, one was a discussion of currently existing methods, twelve were assessments of \textit{learning} (educational) programs that are \textit{targeted} towards clinical environments, and six were comparisons of machine learning algorithms to other prediction models (without mention of TMLE). Overall, we focused on 81 observational studies in this systematic review for which full texts were available; six publications were not open-access and full texts were obtained from the corresponding authors.

\subsection{Dissemination and uptake of the TMLE framework} 

    There has been a growing uptake of the TMLE framework over time, with five or fewer applied publications per year until 2017, and up to 21 in 2021. The majority (58, 71.6\%) of publications using TMLE were published in the last three and a half years, starting in 2019. Most studies (77, 95.1\%) cited the authors of particular TMLE methods they apply, whereas four (4.9\%) did not cite any TMLE references. The large majority of these first epidemiological studies benefit from the expert knowledge of an author who is (or was) part of Professor Mark van der Laan's lab. (Table~\ref{tab:Table2}) \\
    
\begin{landscape}
        \begin{figure}
        \centering
        \includegraphics[width=1.3\textwidth]{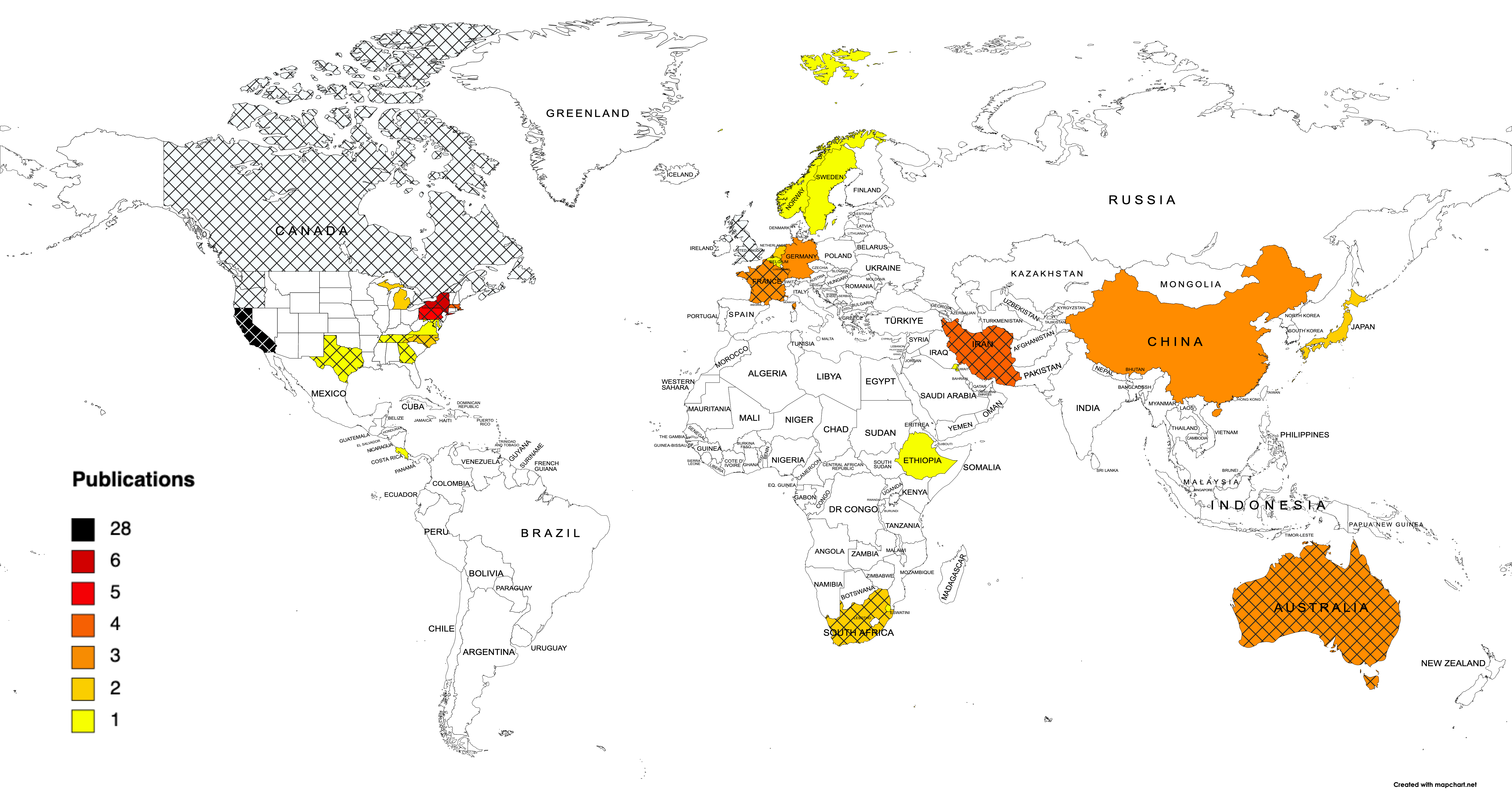}
        \caption{World map of publications using targeted maximum likelihood estimation by the geographical location of the first author (2006 to mid-2022). Colours represent the number of observational studies and the crosshatch pattern identifies where at least one methodological publication stem from.}
        \label{fig:figure2}
        \end{figure}
    \end{landscape}

    Of the 81 studies included, two-thirds were conducted in the United States of America (US) (53, 65.4\%, Figure~\ref{fig:figure2}),\cite{Bembom2009,Rosenblum2009,Mackey2011,Padula2012,Leslie2014,Schnitzer2014,Brown2015,Gianfrancesco2016,Hsu2016,Ahern2016,Salihu2016,Tran2016,Davis2017,Skeem2017,Sukul2017,Akosile2018ReassessingEstimation,Platt2018,Pearl2018,Skeem2018,Mosconi2018,Torres2018,Izano2019,Yu2019,Rudolph2019,Lim2019,Berkowitz2019,Gianfrancesco2019,Casey2019,Torres2019,You2019ApplicationMexico,Izano2020,Ehrlich2020,Bodnar2020,Kagawa2020,Kempker2020,Yu2020,Puryear2020,Torres2020,Westling2020,Aiemjoy2020,Reilly2021,Kang2021,Torres2021a,Torres2021b,Mehta2021,Nardone2021,Lee2021,Shiba2021,Goel2021,Kahkoska2021,Beydoun2021,Crowner2022,Wong2022} with 100\% of articles before 2017 being published in the US and down to 50\% of all 2021 articles. Publications from Europe or Scandinavia (10, 12.3\%),\cite{Legrand2013,Herrera2017,Rodriguez2019,Papadopoulou2019,Vauchel2019ImpactAnalysis,Veit2020,Decruyenaere2020,Rossides2021,Zou2021,Isfordink2022} Africa (4, 4.9\%),\cite{Bell-Gorrod2020,Amusa2021,Kerschberger2021,Dadi2021} Middle East (5, 6.2\%),\cite{Mozafar2020,Abdollahpour2021,Mozafar2021,Almasi2021,Akhtar2022} and Oceania or Asia (8, 9.9\%)\cite{Clare2020,Wang2021,Sun2021,Chavda2022,Chen2022,Ikeda2022a,Ikeda2022b,MorenoBetancur2022} represent between 20\% (in 2017) and 70\% (in 2022, up to May 2022) of all applied studies published in the last 6 years. (Figure \ref{fig:figure2}, Table \ref{tab:Table2}) In the US, the majority of publications (28) were from California, including 20 from the University of California at Berkeley, where TMLE was first described.\\
    
    In the early years, the first authors tended to be qualitative academic experts, but we saw more variety in expertise and a larger number of practising clinicians leading observational studies in many epidemiological and public health fields in recent publications. 
    The most common epidemiological sub-discipline was non-communicable disease (23, 28.4\%),\cite{Legrand2013,Leslie2014,Akosile2018ReassessingEstimation,Mosconi2018,Torres2018,Yu2019,Lim2019,Gianfrancesco2019,Mozafar2020,Veit2020,Yu2020,Decruyenaere2020,Abdollahpour2021,Reilly2021,Torres2021b,Mozafar2021,Almasi2021,Dadi2021,Zou2021,Goel2021,Beydoun2021,Chavda2022,Crowner2022} followed by behavioural epidemiology (17, 21.0\%),\cite{Mackey2011,Ahern2016,Platt2018,Rodriguez2019,Torres2019,Ehrlich2020,Bodnar2020,Kagawa2020,Puryear2020,Torres2020,Clare2020,Kang2021,Torres2021a,Lee2021,Shiba2021,Ikeda2022a,Ikeda2022b} and then infectious disease epidemiology (12, 14.8\%).\cite{Schnitzer2014,Davis2017,Vauchel2019ImpactAnalysis,Figueroa2020EarlyApproaches,Kempker2020,Westling2020,Aiemjoy2020,Amusa2021,Kerschberger2021,Akhtar2022,Chen2022,Isfordink2022} Through time we see an uptake of TMLE in many more disciplines, such as pharmaco-epidemiology,\cite{Sukul2017,Bell-Gorrod2020,Rossides2021,Kahkoska2021} policy,\cite{Tran2016,Skeem2017,Skeem2018,You2019ApplicationMexico,Mehta2021,Wong2022,MorenoBetancur2022} biomarker epidemiology,\cite{Bembom2009,Rosenblum2009,Gianfrancesco2016,Hsu2016,Salihu2016,Izano2020,Wang2021,Sun2021} environmental epidemiology,\cite{Padula2012,Herrera2017,Pearl2018,Rudolph2019,Casey2019,Papadopoulou2019,Nardone2021} occupational epidemiology,\cite{Brown2015,Izano2019} and health economy.\cite{Berkowitz2019} \\

    We also studied the evolution of citations. When only methodological overviews of the TMLE framework were available, these were cited despite their heavy statistical requisite. Since 2016, tutorials were published and started to be cited alongside references for statistical packages. \cite{Gruber2015a,Pang2016a,Pearl2016,Schuler2017,Kreif2017,Luque-Fernandez2018,Almasi2018,Smith2022}(Table \ref{tab:Table2}) \\

    Cohort study design\cite{Rosenblum2009,Mackey2011,Schnitzer2014,Brown2015,Gianfrancesco2016,Skeem2017,Sukul2017,Akosile2018ReassessingEstimation,Skeem2018,Torres2018,Izano2019,Lim2019,Gianfrancesco2019,Casey2019,Torres2019,Vauchel2019ImpactAnalysis,You2019ApplicationMexico,Veit2020,Ehrlich2020,Kempker2020,Torres2020,Westling2020,Clare2020,Bell-Gorrod2020,Reilly2021,Wang2021,Torres2021b,Kerschberger2021,Lee2021,Shiba2021,Dadi2021,Zou2021,Sun2021,Goel2021,Kahkoska2021,Chavda2022,Chen2022,Crowner2022,Isfordink2022,Ikeda2022a,Ikeda2022b,MorenoBetancur2022} was the most prevalent design (42, 51.9\%), followed by cross-sectional (32, 39.5\%) (Appendix Table \ref{table:discipline}).\cite{Bembom2009,Padula2012,Legrand2013,Leslie2014,Ahern2016,Salihu2016,Tran2016,Davis2017,Herrera2017,Platt2018,Pearl2018,Mosconi2018,Yu2019,Rudolph2019,Berkowitz2019,Rodriguez2019,Papadopoulou2019,Izano2020,Mozafar2020,Bodnar2020,Kagawa2020,Yu2020,Puryear2020,Decruyenaere2020,Aiemjoy2020,Amusa2021,Kang2021,Torres2021a,Mozafar2021,Mehta2021,Rossides2021,Beydoun2021} Other types of epidemiological design included case-control\cite{Hsu2016,Figueroa2020EarlyApproaches,Abdollahpour2021,Almasi2021,Akhtar2022} and ecological.\cite{Nardone2021,Wong2022}  \\

    Many articles reported results from other statistical methods, in addition to reporting those obtained from TMLE. Over one-quarter of the studies used adjusted parametric regression (24, 29.6\%),\cite{Padula2012,Leslie2014,Schnitzer2014,Tran2016,Sukul2017,Mosconi2018,Izano2019,Lim2019,Rodriguez2019,Gianfrancesco2019,Papadopoulou2019,Izano2020,Mozafar2020,Bodnar2020,Figueroa2020EarlyApproaches,Westling2020,Decruyenaere2020,Torres2021b,Kerschberger2021,Rossides2021,Dadi2021,Zou2021,Chavda2022,Akhtar2022} one sixth (12, 14.8\%) used IPTW,\cite{Mackey2011,Schnitzer2014,Pearl2018,Yu2019,Ehrlich2020,Bell-Gorrod2020,Amusa2021,Kang2021,Torres2021a,Nardone2021,Shiba2021,Kahkoska2021} one (1.2\%) used AIPTW,\cite{Mehta2021} three (3.7\%) used non-parametric methods (e.g. Kaplan Meier),\cite{Schnitzer2014,Izano2019,Torres2021b} and seven (8.6\%) used unadjusted regression.\cite{Ehrlich2020,Puryear2020,Reilly2021,Crowner2022,Wong2022} Some studies included more than one comparative method.\\
    
    SuperLearner (SL) provides a flexible machine learning approach to  the estimation of the initial outcome and intervention models. Of the 81 articles, more than half (47, 58.0\%) used the SL algorithm, \cite{Tran2016,Herrera2017,Skeem2017,Akosile2018ReassessingEstimation,Pearl2018,Skeem2018,Mosconi2018,Torres2018,Yu2019,Rodriguez2019,Gianfrancesco2019,Torres2019,Vauchel2019ImpactAnalysis,You2019ApplicationMexico,Izano2020,Ehrlich2020,Figueroa2020EarlyApproaches,Kagawa2020,Kempker2020,Yu2020,Torres2020,Westling2020,Decruyenaere2020,Clare2020,Aiemjoy2020,Amusa2021,Reilly2021,Wang2021,Torres2021a,Torres2021b,Mozafar2021,Mehta2021,Almasi2021,Kerschberger2021,Nardone2021,Rossides2021,Shiba2021,Sun2021,Kahkoska2021,Beydoun2021,Chavda2022,Akhtar2022,Chen2022,Crowner2022,Wong2022,Isfordink2022,Ikeda2022a,Ikeda2022b,MorenoBetancur2022} 18 (22.2\%) used logistic regression,\cite{Bembom2009,Rosenblum2009,Mackey2011,Padula2012,Leslie2014,Schnitzer2014,Brown2015,Ahern2016,Sukul2017,Platt2018,Izano2019,Rudolph2019,Lim2019,Bodnar2020,Bell-Gorrod2020,Abdollahpour2021,Kang2021} and 16 (19.8\%) did not specify the approach for the estimation of either the outcome or intervention model.\cite{Legrand2013,Gianfrancesco2016,Hsu2016,Salihu2016,Davis2017,Berkowitz2019,Casey2019,Papadopoulou2019,Mozafar2020,Veit2020,Puryear2020,Lee2021,Dadi2021,Zou2021,Goel2021} The average number of machine-learning algorithms considered by the SL was 6.3 (range 1 - 16), 19 different machine-learning algorithms were used across the articles (a machine-learning algorithm is a wrapper included within the SuperLearner\cite{vanderLaan2011TargetedLearning} library in R software).\\
    
    Variance of point estimates obtained from TMLE were estimated using the influence function (n=19, 23.5\%),\cite{Padula2012,Legrand2013,Leslie2014,Schnitzer2014,Brown2015,Hsu2016,Tran2016,Platt2018,Lim2019,Mozafar2020,Decruyenaere2020,Aiemjoy2020,Abdollahpour2021,Mozafar2021,Almasi2021,Rossides2021,Chavda2022,Akhtar2022,Wong2022} bootstrap (n=6, 7.4\%),\cite{Rosenblum2009,Mackey2011,Sukul2017,Pearl2018,Yu2019,Kang2021} and Wald tests (n=2, 2.5\%),\cite{Ahern2016,Skeem2017} while 54 (66.7\%) studies did not specify how standard errors were obtained.\cite{Bembom2009,Gianfrancesco2016,Salihu2016,Davis2017,Herrera2017,Skeem2018ComparingIllness,Mosconi2018,Torres2018,Akosile2018ReassessingEstimation,Izano2019,Rudolph2019,Berkowitz2019,Rodriguez2019,Gianfrancesco2019,Casey2019,Papadopoulou2019,Torres2019,Vauchel2019ImpactAnalysis,You2019ApplicationMexico,Figueroa2020EarlyApproaches,Izano2020,Veit2020,Ehrlich2020,Bodnar2020,Kagawa2020,Kempker2020,Yu2020,Puryear2020,Torres2020,Westling2020,Clare2020,Bell-Gorrod2020,Amusa2021,Reilly2021,Wang2021,Torres2021a,Torres2021b,Mehta2021,Kerschberger2021,Nardone2021,Lee2021,Shiba2021,Dadi2021,Zou2021,Sun2021,Goel2021,Kahkoska2021,Beydoun2021,Chen2022,Crowner2022,Isfordink2022,Ikeda2022a,Ikeda2022b,MorenoBetancur2022}\\
    
    The Causal Inference Roadmap\cite{Saddiki2018AScience} contains seven recommended criteria to define a causal effect: (i) specify the scientific question, (ii) specify the causal model, (iii) define the target causal quantity, (iv) link the observed data to the causal model, (v) assess identifiability assumptions, (vi) estimate the target statistical parameters, and (vii) interpretation of the results. On average, 5.3 (SD 1.0) criteria were complete per article. We considered a version of the Targeted Learning Roadmap\cite{Coyle2020TargetingResearch} that contains five criteria: (i) specify the observed data and describe the data-generating experiment, (ii) specify a statistical model representing a set of realistic assumptions about the underlying true probability distribution of the data, (iii) define a target estimand of the data distribution that ``best" approximates the answer to the scientific question of interest, (iv) given statistical model and target estimand, construct an optimal plug-in estimator of the target estimand of the observed data distribution, while respecting the model, and (v) construct a confidence interval by estimating the sampling distribution of the estimator. When the scientific question of interest is causal, step (iii) of the Targeted Learning Roadmap incorporates steps (ii)--(v) of the Causal Inference Roadmap.\cite{Coyle2020TargetingResearch} On average, 3.3 (SD 1.0) criteria were complete per article. Most studies have room to state the necessary content for at least one more criteria. \\
    
    Most publications (77, 95.1\%) used R software to perform TMLE,\cite{Mackey2011,Padula2012,Legrand2013,Leslie2014,Schnitzer2014,Brown2015,Gianfrancesco2016,Hsu2016,Ahern2016,Tran2016,Davis2017,Herrera2017,Skeem2017,Sukul2017,Akosile2018ReassessingEstimation,Platt2018,Pearl2018,Skeem2018,Mosconi2018,Torres2018,Izano2019,Yu2019,Lim2019,Berkowitz2019,Rodriguez2019,Gianfrancesco2019,Casey2019,Torres2019,Vauchel2019ImpactAnalysis,You2019ApplicationMexico,Izano2020,Mozafar2020,Veit2020,Ehrlich2020,Bodnar2020,Kagawa2020,Kempker2020,Yu2020,Puryear2020,Torres2020,Decruyenaere2020,Clare2020,Aiemjoy2020,Bell-Gorrod2020,Figueroa2020EarlyApproaches,Amusa2021,Abdollahpour2021,Reilly2021,Wang2021,Torres2021a,Torres2021b,Mozafar2021,Mehta2021,Almasi2021,Kerschberger2021,Nardone2021,Lee2021,Rossides2021,Zou2021,Sun2021,Akhtar2022,Crowner2022,Wong2022,Isfordink2022,Ikeda2022b,MorenoBetancur2022} except four that used STATA.\cite{Papadopoulou2019,Dadi2021,Beydoun2021,Chavda2022} Nonetheless, nine articles reported using another software tool (i.e., Stata/SAS/SPSS) alongside R for TMLE.\cite{Mackey2011,Mosconi2018,Berkowitz2019,Kagawa2020,Bell-Gorrod2020,Almasi2021,Kerschberger2021,Rossides2021,Crowner2022} The most commonly used R software packages were \textit{tmle}\cite{Gruber2012Tmle:Estimation} (40, 49.4\%) and \textit{ltmle}\cite{Lendle2017Ltmle:Data} (16, 19.8\%).

\subsection{Showcase of the TMLE framework in different disciplines} 
    
    In all disciplines and applications, applying the TMLE framework to their specific research question encouraged authors to review the strengths and limitations of their data and carefully consider how their data and setting might violate identifiability assumptions, which are assumptions necessary for causal inference but not TMLE. If, and only if, the identifiability assumptions are assumed to hold, the estimated effect is a causal effect. However, for observational studies, it cannot be known whether identifiability assumptions hold. Therefore, if an estimate is interpreted as a causal effect, then this interpretation should be accompanied by a discussion of the plausibility of identifiability assumptions. All disciplines and disease areas highlight issues with missing data and measurement errors and incorporate subject-matter knowledge. (Appendix Table \ref{table:discipline}) Model misspecification, which might result from imposing constraints that are unrealistic or not informed by subject-matter knowledge, was the first and foremost driver for using the TMLE framework. Three-quarters of the studies (n=68, 84.0\%) provided at least one justification for using TMLE compared to another method. (Table \ref{tab:Table2}) Standard regression techniques in settings with low incidence,\cite{Davis2017} rare outcomes\cite{Bell-Gorrod2020} or low sample size\cite{Rodriguez2019, Kang2021} may over-fit the data or not converge: careful SL specifications overcome these limitations.\cite{Davis2017,Berkowitz2019} Least biased and double-robust estimation were also widely cited advantages of the framework. These mean (i) the estimated parameter will be closest to the true value of our quantity of interest, even when our sample size is low, and (ii) only one of our initial models needs to be correctly specified.   \\
    
    There was a range of disease areas covered in the 23 \textbf{noncommunicable disease epidemiology} studies. The appealing property of TMLE was that it is a semiparametric estimator, allowing the use of machine learning algorithms to minimise model misspecification.\cite{Beydoun2021,Yu2019,Goel2021,Yu2020,Mozafar2021,Dadi2021,Chavda2022,Zou2021} Additionally, extensions of TMLE have developed ways to appropriately handle the dual nature of time-varying confounding, which have been utilised in longitudinal studies analysing data on depression,\cite{Torres2018} survival from acute respiratory distress syndrome,\cite{Torres2021b} caries arising from intake of soda,\cite{Lim2019} effects of smoking on rheumatoid arthritis,\cite{Gianfrancesco2019} and effects of asthma medication on asthma symptoms.\cite{Veit2020} Improved predictive performance\cite{Legrand2013} and adjusting for informative censoring\cite{Torres2021b} were additional reasons for using TMLE. Furthermore, the extension of TMLE to case-control studies, in which sampling is biased with respect to the disease status, provided a platform for analysing the causal effect of reproductive factors on breast cancer by using case-control weighted TMLE.\cite{Almasi2021} \\
    
    In \textbf{infectious disease epidemiology} (IDE) articles, most were concerned with having a flexible modelling approach that does not impose assumptions on the functional form of the exposure-outcome relationship.\cite{Aiemjoy2020,Amusa2021,Chen2022,Vauchel2019ImpactAnalysis,Figueroa2020EarlyApproaches,Kempker2020} A key feature of the IDE subdiscipline is that baseline confounders and exposures may change over time and can obscure the causal effect of interest.\cite{Schnitzer2014} Standard survival modelling assumes that censoring and survival processes are independent, which is likely violated in this setting, and it assumes there is no time-dependent confounding.\cite{Schnitzer2014} TMLEs with a working marginal structural model and for time-to-event outcomes permit evaluation of the effect of an exposure at multiple time points, which is beneficial when the interpretation of causal effects from hazard models is often difficult.\cite{Hernan2010TheRatios} Other studies have overcome this issue by using TMLE in a target trial framework or case-cohort studies.\cite{Westling2020,Akhtar2022} \\ 
    
    In \textbf{behavioural epidemiology} manuscripts, the behavioural nature of the topics covered implied that RCTs are mostly unethical, bear prohibitive costs or have very small sample sizes. There are several key challenges for using observational data to study the causal effects of childhood adversities,\cite{Ahern2016,Platt2018} physical activity,\cite{Mackey2011,Ehrlich2020, Ikeda2022b} alcohol consumption\cite{Kagawa2020} or supply \cite{Clare2020} on various outcomes, including fractures,\cite{Mackey2011} mental health,\cite{Ahern2016,Lee2021,Shiba2021} asthma,\cite{Rodriguez2019} and pregnancy outcomes.\cite{Ehrlich2020,Bodnar2020} They include a risk for reverse causation;\cite{Torres2021b,Lee2021,Shiba2021} high dimensional data and in particular, multidimensional exposures;\cite{Ahern2016,Platt2018} and measurement error resulting from self-reported exposures or outcomes.\cite{Bodnar2020,Shiba2021,Ikeda2022a,Ikeda2022b} Longitudinal relationships and time-varying confounding, where confounders of the effect of an exposure on an outcome can themselves be affected by prior exposures, as well as sample attrition,\cite{Torres2019,Clare2020,Ikeda2022a,Ikeda2022b} are particular challenges faced by survey data that are collected in consecutive waves.\cite{Torres2019,Clare2020,Torres2021a,Shiba2021,Ikeda2022a,Ikeda2022b} TMLE adjusts for time-varying confounders affected by prior exposure and employs a doubly robust estimation approach that allows for flexible model fitting. Additionally, as pointed out in 2016 by Ahern et al.,\cite{Ahern2016} ``TMLE with machine learning addresses the challenge of a multidimensional exposure because it facilitates `learning' from the data the strength of the relations between each adversity [dimensions of the exposure] and outcome, incorporating any interactions or nonlinearity, specific to each [sub-group]". \\ 
    
    The field of \textbf{biomarker epidemiology} is driven by the search for sets of candidate biomarkers that are important in determining given outcomes. Ranking the contributions of these candidate biomarkers is also of interest. Some studies used TMLE to measure variable importance in biomarker research\cite{Bembom2009,Hsu2016,Sun2021} and in other fields.\cite{Legrand2013} Dimension reduction for the estimation of causal effects is an aim in some biomarker examples.\cite{Salihu2016,Izano2020,Wang2021} In the models presented in the publications, there are complex joint effects to consider in large correlated data, as well as longitudinal patterns and time-dependent confounding.\cite{Salihu2016,Izano2020,Wang2021} Furthermore, two manuscripts present covariate selection algorithms ahead of causal effect estimation.\cite{Wang2021,Sun2021} \\
    
    Research published in \textbf{environmental epidemiology} highlights challenges around the clear definitions of exposure and outcomes,\cite{Padula2012} as there are likely many proxy and surrogate measures of exposure,\cite{Herrera2017} coupled with potential exposure misclassification and measurement errors.\cite{Padula2012,Rudolph2019} Nonetheless, TMLE was successfully applied to determine either causal attributable risk,\cite{Padula2012,Herrera2017} or risk differences.\cite{Pearl2018} \\
    
    The only observational study of TMLE in \textbf{health economics} explored the relationship between financial resources leading to food insecurity and healthcare expenditure in a pay-to-access healthcare system. It uses ecological measures of exposure and outcome and leads to evidence for policy.\cite{Berkowitz2019} \\ 
    
    Two publications focused on \textbf{occupational epidemiology}.\cite{Brown2015,Izano2019} A key aspect of occupational epidemiology is accounting for the healthy worker survivor effect: a bias arising due to healthier workers accruing more exposure over time. These studies looked at exposure to particulate matter from aluminium or metalworking fluids in metal factory workers, which varied depending on the length of employment. Both studies benefited from TMLE's flexibility to allow for time-varying confounding of the exposure.\\

    The field of \textbf{pharmacoepidemiology} is concerned with assessing treatment's efficacy in real-world settings and monitoring long-term side effects of treatments. Both objectives would be either impractical or too costly to study in RCTs, given the limited follow-up time available in clinical trials. TMLE has been used in this setting, as it provides a robust procedure for estimation.\cite{Sukul2017,Bell-Gorrod2020,Rossides2021,Kahkoska2021} In particular, the flexibility of TMLE, provided through the specification of a diverse and rich set of machine learning algorithms in the SL, is crucial for appropriately adjusting for confounding in observational studies.\cite{Phillips2022PracticalLearner} \\ 

    \textbf{Policy} epidemiology assesses the effects of population programs or mandates. Lack of randomisation, such as in studies examining the association between speciality probation and public safety outcomes,\cite{Skeem2017,Skeem2018} leads to an imbalance in the covariate distribution by exposure levels. Studies of cost-effectiveness may involve dealing with outliers which can be addressed with TMLE.\cite{Skeem2018,Mehta2021} Other challenges include zero-inflation, such as the assessment of the effect of primary care physician density on arthroplasty outcomes, in which some areas had zero density.\cite{Mehta2021} This is dealt with by using a mixture of models to assess the probability of non-exposure (i.e., very low density).\cite{Mehta2021} Other policy studies presented challenges around missing data,\cite{Mehta2021} reliance on epidemic modelling assumptions,\cite{Wong2022} target trial emulation,\cite{MorenoBetancur2022} or infeasible randomisation process.\cite{You2019ApplicationMexico} 
    
    \newpage 
    
    \begin{landscape}
        \resizebox{23cm}{!}{
        \begin{threeparttable}[ht]
        \centering
        \caption{Distribution of observational papers by year of publication and selected characteristics}
        \label{tab:Table2}
        \begin{tabular}{r >{\bfseries}r r >{\bfseries}r r >{\bfseries}r r >{\bfseries}r r >{\bfseries}r r >{\bfseries}r r >{\bfseries}r r >{\bfseries}r r >{\bfseries}r r >{\bfseries}r r >{\bfseries}r r >{\bfseries}r r >{\bfseries}r r >{\bfseries}r r >{\bfseries}r}
        \multicolumn{1}{l}{} & \multicolumn{28}{c}{\textbf{Year of publication}} & \multicolumn{1}{l}{} \\ \cline{2-29}
         & \multicolumn{2}{c}{\textbf{2009}} & \multicolumn{2}{c}{\textbf{2010}} & \multicolumn{2}{c}{\textbf{2011}} & \multicolumn{2}{c}{\textbf{2012}} & \multicolumn{2}{c}{\textbf{2013}} & \multicolumn{2}{c}{\textbf{2014}} & \multicolumn{2}{c}{\textbf{2015}} & \multicolumn{2}{c}{\textbf{2016}} & \multicolumn{2}{c}{\textbf{2017}} & \multicolumn{2}{c}{\textbf{2018}} & \multicolumn{2}{c}{\textbf{2019}} & \multicolumn{2}{c}{\textbf{2020}} & \multicolumn{2}{c}{\textbf{2021}} & \multicolumn{2}{c}{\textbf{2022 $^{\star}$}} & \textbf{Total} \\ \cline{2-30} 
         & N & (\%) & N & (\%) & N & (\%) & N & (\%) & N & (\%) & N & (\%) & N & (\%) & N & (\%) & N & (\%) & N & (\%) & N & (\%) & N & (\%) & N & (\%) & N & (\%) & N \\ \cline{2-30}
        \multicolumn{1}{l}{} & \multicolumn{1}{l}{} & \multicolumn{1}{l}{} & \multicolumn{1}{l}{} & \multicolumn{1}{l}{} & \multicolumn{1}{l}{} & \multicolumn{1}{l}{} & \multicolumn{1}{l}{} & \multicolumn{1}{l}{} & \multicolumn{1}{l}{} & \multicolumn{1}{l}{} & \multicolumn{1}{l}{} & \multicolumn{1}{l}{} & \multicolumn{1}{l}{} & \multicolumn{1}{l}{} & \multicolumn{1}{l}{} & \multicolumn{1}{l}{} & \multicolumn{1}{l}{} & \multicolumn{1}{l}{} & \multicolumn{1}{l}{} & \multicolumn{1}{l}{} & \multicolumn{1}{l}{} & \multicolumn{1}{l}{} & \multicolumn{1}{l}{} & \multicolumn{1}{l}{} & \multicolumn{1}{l}{} & \multicolumn{1}{l}{} & \multicolumn{1}{l}{} & \multicolumn{1}{l}{} & \multicolumn{1}{l}{} \\
        \multicolumn{1}{l}{\textbf{Publications}} & 2 &  &  &  & 1 &  & 1 &  & 1 &  & 2 &  & 1 &  & 5 &  & 4 &  & 6 &  & 12 &  & 16 &  & 21 &  & 9 &  & 81 \\ 
        \multicolumn{1}{l}{} & \multicolumn{1}{l}{} & \multicolumn{1}{l}{} & \multicolumn{1}{l}{} & \multicolumn{1}{l}{} & \multicolumn{1}{l}{} & \multicolumn{1}{l}{} & \multicolumn{1}{l}{} & \multicolumn{1}{l}{} & \multicolumn{1}{l}{} & \multicolumn{1}{l}{} & \multicolumn{1}{l}{} & \multicolumn{1}{l}{} & \multicolumn{1}{l}{} & \multicolumn{1}{l}{} & \multicolumn{1}{l}{} & \multicolumn{1}{l}{} & \multicolumn{1}{l}{} & \multicolumn{1}{l}{} & \multicolumn{1}{l}{} & \multicolumn{1}{l}{} & \multicolumn{1}{l}{} & \multicolumn{1}{l}{} & \multicolumn{1}{l}{} & \multicolumn{1}{l}{} & \multicolumn{1}{l}{} & \multicolumn{1}{l}{} & \multicolumn{1}{l}{} & \multicolumn{1}{l}{} & \multicolumn{1}{l}{} \\ \arrayrulecolor{gray}\hline
        \multicolumn{1}{l}{\textbf{TMLE expert (author)$^{\pm}$ }} & 2 & (100) &  &  & 1 & (100) & 1 & (100) & 1 & (100) & 1 & (50) & 1 & (100) & 3 & (60) & 1 & (25) & 4 & (67) & 6 & (50) & 6 & (38) & 5 & (24) & 1 & (11) & 33 \\
        \multicolumn{1}{l}{} & \multicolumn{1}{l}{} & \multicolumn{1}{l}{} & \multicolumn{1}{l}{} & \multicolumn{1}{l}{} & \multicolumn{1}{l}{} & \multicolumn{1}{l}{} & \multicolumn{1}{l}{} & \multicolumn{1}{l}{} & \multicolumn{1}{l}{} & \multicolumn{1}{l}{} & \multicolumn{1}{l}{} & \multicolumn{1}{l}{} & \multicolumn{1}{l}{} & \multicolumn{1}{l}{} & \multicolumn{1}{l}{} & \multicolumn{1}{l}{} & \multicolumn{1}{l}{} & \multicolumn{1}{l}{} & \multicolumn{1}{l}{} & \multicolumn{1}{l}{} & \multicolumn{1}{l}{} & \multicolumn{1}{l}{} & \multicolumn{1}{l}{} & \multicolumn{1}{l}{} & \multicolumn{1}{l}{} & \multicolumn{1}{l}{} & \multicolumn{1}{l}{} & \multicolumn{1}{l}{} & \multicolumn{1}{l}{} \\  \arrayrulecolor{gray}\hline
        \multicolumn{1}{l}{\textbf{USA-based publication}} & 2 & (100) &  &  & 1 & (100) & 1 & (100) &  &  & 2 & (100) & 1 & (100) & 5 & (100) & 3 & (75) & 6 & (100) & 9 & (75) & 10 & (63) & 11 & (52) & 2 & (22) & 52 \\ 
        \multicolumn{1}{l}{} &  &  &  &  &  &  &  &  &  &  &  &  &  &  &  &  &  &  &  &  &  &  &  &  &  &  &  &  &  \\ \arrayrulecolor{gray}\hline
        \multicolumn{1}{l}{\textbf{Discipline}} &  &  &  &  &  &  &  &  &  &  &  &  &  &  &  &  &  &  &  &  &  &  &  &  &  &  &  &  &  \\
        \textit{Behavioural epi} &  &  &  &  & 1 & (100) &  &  &  &  &  &  &  &  & 1 & (20) &  &  & 1 & (17) & 2 & (17) & 6 & (38) & 4 & (19) & 2 & (22) & 17 \\
        \textit{Biomarker} & 2 & (100) &  &  &  &  &  &  &  &  &  &  &  &  & 3 & (60) &  &  &  &  &  &  & 1 & (6) & 2 & (10) &  &  & 8 \\
        \textit{Environmental epi} &  &  &  &  &  &  & 1 & (100) &  &  &  &  &  &  &  &  & 1 & (25) & 1 & (17) & 3 & (25) &  &  & 1 & (5) &  &  & 7 \\
        \textit{Health economy} &  &  &  &  &  &  &  &  &  &  &  &  &  &  &  &  &  &  &  &  & 1 & (8) &  &  &  &  &  &  & 1 \\
        \textit{Infectious disease} &  &  &  &  &  &  &  &  &  &  & 1 & (50) &  &  &  &  & 1 & (25) &  &  & 1 & (8) & 4 & (25) & 2 & (10) & 3 & (33) & 12 \\
        \textit{Non-Communicable Disease} &  &  &  &  &  &  &  &  & 1 & (100) & 1 & (50) &  &  &  &  &  &  & 3 & (50) & 3 & (25) & 4 & (25) & 9 & (43) & 2 & (22) & 23 \\
        \textit{Occupational epi} &  &  &  &  &  &  &  &  &  &  &  &  & 1 & (100) &  &  &  &  &  &  & 1 & (8) &  &  &  &  &  &  & 2 \\
        \textit{Pharmaco-epi} &  &  &  &  &  &  &  &  &  &  &  &  &  &  &  &  & 1 & (25) &  &  &  &  & 1 & (6) & 2 & (10) &  &  & 4 \\
        \textit{Policy} &  &  &  &  &  &  &  &  &  &  &  &  &  &  & 1 & (20) & 1 & (25) & 1 & (17) & 1 & (8) &  &  & 1 & (5) & 2 & (22) & 7 \\
        \multicolumn{1}{l}{} &  &  &  &  &  &  &  &  &  &  &  &  &  &  &  &  &  &  &  &  &  &  &  &  &  &  &  &  &  \\ \arrayrulecolor{gray}\hline
        \multicolumn{1}{l}{\textbf{Motivations $^{\dagger}$}} &  &  &  &  &  &  &  &  &  &  &  &  &  &  &  &  &  &  &  &  &  &  &  &  &  &  &  &  &  \\
        \textit{Bias} & 1 & (50) &  &  &  &  &  &  &  &  & 2 & (100) & 1 & (100) & 1 & (20) & 1 & (25) & 2 & (33) & 5 & (42) & 5 & (31) & 12 & (57) & 6 & (67) & 36 \\
        \textit{Double-robust} & 1 & (50) &  &  &  &  &  &  &  &  &  &  &  &  & 1 & (20) & 1 & (25) & 2 & (33) & 2 & (17) & 5 & (31) & 6 & (29) & 4 & (44) & 22 \\
        \textit{Efficient} &  &  &  &  &  &  &  &  &  &  &  &  &  &  &  &  &  &  &  &  & 2 & (17) & 5 & (31) & 2 & (10) & 3 & (33) & 12 \\
        \textit{Finite   sample} &  &  &  &  &  &  & 1 & (100) &  &  &  &  &  &  &  &  &  &  &  &  &  &  &  &  & 2 & (10) &  &  & 3 \\
        \textit{Model misspecification} & 1 & (50) &  &  & 1 & (100) & 1 & (100) &  &  &  &  &  &  & 3 & (60) & 1 & (25) & 2 & (33) & 1 & (8) & 3 & (17) & 1 & (57) & 2 & (22) & 16 \\
        \textit{Positivity   assumption} &  &  &  &  &  &  &  &  &  &  &  &  &  &  &  &  &  &  & 1 & (17) & 1 & (8) & 1 & (6) &  &  &  &  & 3 \\
        \textit{Power} &  &  &  &  &  &  &  &  &  &  &  &  &  &  &  &  &  &  &  &  &  &  & 1 & (6) &  &  &  &  & 1 \\
        \textit{None specified} &  &  &  &  &  &  &  &  & 1 & (100) &  &  &  &  & 1 & (20) & 2 & (50) &  &  & 5 & (42) & 4 & (25) & 2 & (10) & 1 & (11) & 16 \\
        \multicolumn{1}{l}{} &  &  &  &  &  &  &  &  &  &  &  &  &  &  &  &  &  &  &  &  &  &  &  &  &  &  &  &  &  \\ \arrayrulecolor{gray}\hline
        \multicolumn{1}{l}{\textbf{Expertise first author}} &  &  &  &  &  &  &  &  &  &  &  &  &  &  &  &  &  &  &  &  &  &  &  &  &  &  &  &  &  \\
        \textit{Biostatistician} & 1 & (50) &  &  &  &  &  &  &  &  & 1 & (50) & 1 & (100) &  &  & 1 & (25) & 2 & (33) & 2 & (17) & 2 & (13) & 4 & (19) & 2 & (22) & 16 \\
        \textit{Epidemiologist} &  &  &  &  & 1 & (100) & 1 & (100) &  &  & 1 & (50) &  &  & 3 & (60) &  &  & 1 & (17) & 4 & (33) & 6 & (38) & 7 & (33) & 1 & (11) & 25 \\
        \textit{MD} &  &  &  &  &  &  &  &  &  &  &  &  &  &  & 1 & (20) & 2 & (50) &  &  & 1 & (8) & 1 & (6) & 5 & (24) & 2 & (22) & 12 \\
        \textit{MD, MPH} &  &  &  &  &  &  &  &  &  &  &  &  &  &  &  &  &  &  &  &  & 1 & (8) & 1 & (6) &  &  &  &  & 2 \\
        \textit{MD, PhD} &  &  &  &  &  &  &  &  & 1 & (100) &  &  &  &  &  &  &  &  &  &  & 3 & (25) & 2 & (13) & 3 & (14) & 3 & (33) & 12 \\
        \textit{Other} &  &  &  &  &  &  &  &  &  &  &  &  &  &  &  &  &  &  &  &  &  &  & 2 & (13) & 2 & (10) &  &  & 4 \\
        \textit{PhD} & 1 & (50) &  &  &  &  &  &  &  &  &  &  &  &  &  &  & 1 & (25) & 2 & (33) &  &  & 1 & (6) &  &  &  &  & 5 \\
        \textit{Not known} &  &  &  &  &  &  &  &  &  &  &  &  &  &  & 1 & (20) &  &  & 1 & (17) & 1 & (8) & 1 & (6) &  &  & 1 & (11) & 5 \\
        \multicolumn{1}{l}{} &  &  &  &  &  &  &  &  &  &  &  &  &  &  &  &  &  &  &  &  &  &  &  &  &  &  &  &  &  \\ \arrayrulecolor{gray}\hline
        \multicolumn{1}{l}{\textbf{Citations $^{\boxdot}$}} &  &  &  &  &  &  &  &  &  &  &  &  &  &  &  &  &  &  &  &  &  &  &  &  &  &  &  &  &  \\
        \textit{Overall TMLE method.} & 2 & (100) &  &  & 1 & (50) & 1 & (100) &  &  & 5 & (56) & 1 & (33) & 4 & (33) & 4 & (50) & 3 & (33) & 6 & (19) & 13 & (42) & 20 & (38) & 4 & (16) & 64 \\
        \textit{Specific TMLE method.} &  &  &  &  & 1 & (50) &  &  & 1 & (100) & 4 & (44) & 1 & (33) & 6 & (50) & 1 & (13) & 3 & (33) & 15 & (47) & 7 & (23) & 13 & (25) & 8 & (32) & 60 \\
        \textit{Tutorial} &  &  &  &  &  &  &  &  &  &  &  &  &  &  &  &  &  &  & 1 & (11) & 7 & (22) & 7 & (23) & 14 & (27) & 7 & (28) & 36 \\
        \textit{R software} &  &  &  &  &  &  &  &  &  &  &  &  & 1 & (33) & 2 & (17) & 3 & (38) & 2 & (22) & 4 & (13) & 4 & (13) & 5 & (10) & 6 & (24) & 27 \\ \hline
        \multicolumn{30}{l}{\multirow{2}{*}{\begin{tabular}[c]{@{}l@{}} $^{\star}$ Up to 20th May 2022 \\
        $^{\pm}$ TMLE expert is a current or past member of M.J. van der Laan's Lab. \\
        $^{\dagger}$ Proportions calculated over the number of publications within that year. \\
        $^{\boxdot}$ Proportions calculated over the total number of citations within that year. \\
        \end{tabular}}} \\
        \end{tabular}
        \end{threeparttable}}
    \end{landscape}
    \newpage

\subsection{Methodological developments and their implementation}
    
    Over the years since the TMLE framework was first laid out,\cite{vanderLaan2006} many contributions have been made to expand the settings in which TMLE is used, provide tools for implementation in standard software, and describe the TMLE framework and application in lay language. Thanks to this, the community of public health researchers and epidemiologists have started implementing the TMLE framework and its latest developments to obtain double robust, least biased and efficient estimates and statistical inference from studies. The properties of TMLE, in contrast to other estimators commonly used for causal inference, include that it is loss-based, well-defined, unbiased, efficient and can be used as a substitution estimator. \\
    
\begin{figure}[ht]
        \centering
        \includegraphics[width=1.0\textwidth]{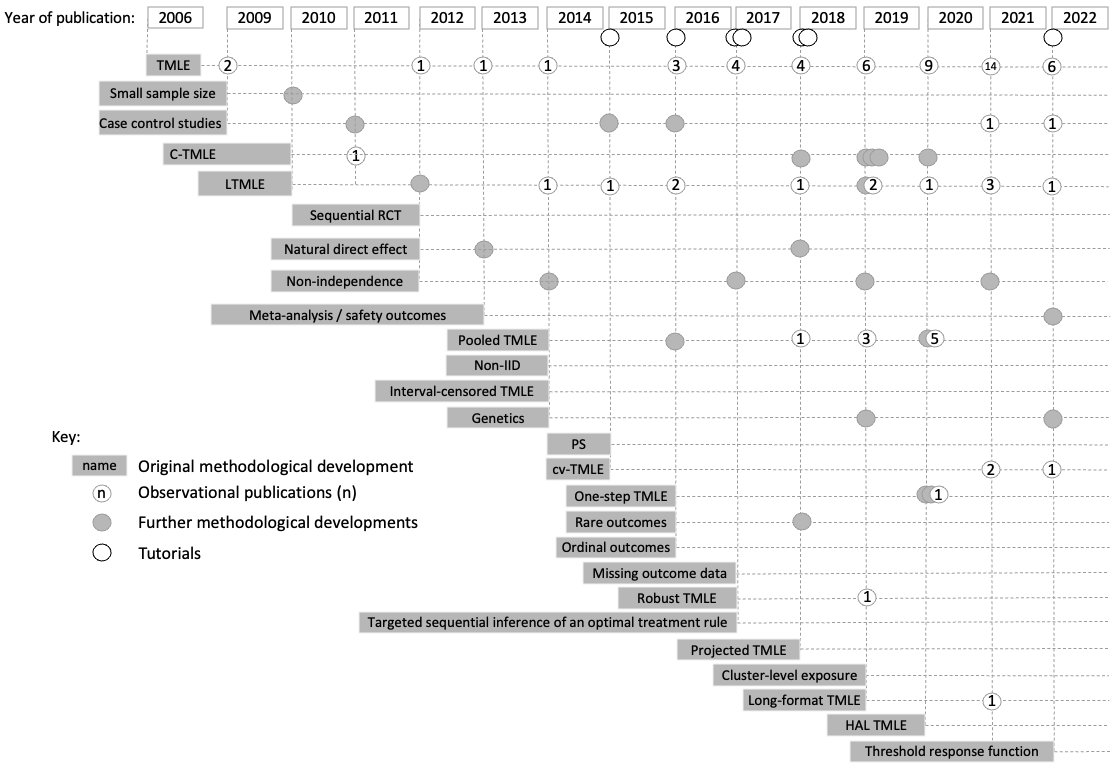}
        \caption{Applied clinical and epidemiological research by year of publication and TMLE method implemented}
        \label{fig:figure3}
    \end{figure}

    Figure \ref{fig:figure3} shows schematically when and why extensions of TMLE have happened in the last 15 years, as well as extensions and uptake. The 81 applied epidemiological studies are classified by methodological development used during the study. Appendix Table \ref{tab:Table3} provides further details regarding the methodological development references between 2006 and mid-2022 that are highlighted in Figure \ref{fig:figure3}. \\
    
    TMLE’s superior efficiency and power are evidenced in small sample size settings where marginal effects from logistic regression models adjusted for (possibly many) covariates would not be recommended.\cite{Moore2009} The implementation of TMLE in complex causal effect estimation problems is discussed in many publications, such as in settings with multiple time point interventions,\cite{vanderLaan2010a,vanderLaan2010b} longitudinal data,\cite{vanderLaan2012a,Schomaker2019} post-intervention effect modifiers,\cite{Zheng2018} dependence of the treatment assignment between units\cite{vanderLaan2012b} or censoring,\cite{Schnitzer2016a} causally connected units,\cite{vanderLaan2014,Sofrygin2017} hierarchical data structures,\cite{Balzer2019} randomisation at the cluster level,\cite{Balzer2021a} large electronic health record data, \cite{Sofrygin2019} and in meta-analyses.\cite{Gruber2013,Liu2022} \\
    
    The TMLE framework is extended and discussed in the setting of case-control studies. One study matched cases to controls,\cite{Rose2009} another used two-stage sampling and nested case-control design.\cite{Rose2011} Other studies required the design to be adaptive to possibly invalid assumptions of independent units\cite{Balzer2015} or if the sample population differs from the (possibly ill-defined) target population.\cite{Balzer2016b} \\
    
    The collaborative TMLE (C-TMLE), introduced in 2010,\cite{vanderLaan2010c} is an extension of TMLE, in which information on the causal parameter of interest is used when estimating and selecting initial model(s). C-TMLE aims to improve the robustness and efficiency of the TMLE. Schnitzer \textit{et al}.\cite{Schnitzer2016a} highlight the pitfalls and their consequences of automated variable selection in causal inference, such as in the propensity score model, and how C-TMLE corrects for this. C-TMLE was later extended to measure variable importance\cite{Pirracchio2018} and to longitudinal data settings.\cite{Schnitzer2020} Proposals to enhance the C-TMLE algorithm include ordering covariates to decrease C-TMLE time complexity,\cite{Ju2019a} using LASSO with C-TMLE for the estimation of the propensity scores,\cite{Ju2019b} and adaptive truncation of the propensity scores with C-TMLE to ensure positivity.\cite{Ju2019c} \\

    The pooled TMLE\cite{Peterson2014} was developed for the context of longitudinal data structures with baseline covariates, time-dependent intervention nodes, intermediate time-dependent covariates, and a possibly time-dependent outcome. Extensions include advice for the optimal discretisation of time\cite{Ferreira2020} and to the hazard function.\cite{Zheng2016}\\

    The one-step TMLE aims to preserve the performance of the original two-step TMLE, and achieves bias reduction in one step (i.e., without additional iterations of the TMLE update step and possible over-fitting in finite samples).\cite{VanDerLaan2016One-StepSubmodels} This one-step TMLE was later extended to counterfactual average survival curves\cite{Cai2020b} and heterogeneous treatment effects.\cite{Zhu2020}\\

    The robust TMLE was proposed in 2017 for transporting intervention effects from one population to another.\cite{Rudolph2017}\\

    The cross-validated TMLE (CV-TMLE) provides asymptotic inference under minimal conditions (i.e., non-parametric smoothness\cite{vanderLaan2015}) keeping the bounds of the parameter estimates. It is also used in the estimation of data-adaptive target parameters, like optimal treatment regimes. Recently, TMLE was shown to be useful in defining thresholds and marking specified levels of risks.\cite{vanderLaan2022} \\
    
    The set of observational articles that use TMLE in their main or sensitivity analyses shows that TMLE has successfully been used to examine associations,\cite{Bembom2009,Mackey2011,Legrand2013,Hsu2016,Ahern2016,Salihu2016,Davis2017,Sukul2017,Platt2018,Pearl2018,Mosconi2018,Torres2018,Yu2019,Rudolph2019,Berkowitz2019,Casey2019,Torres2019,Izano2020,Mozafar2020,Ehrlich2020,Bodnar2020,Kempker2020,Yu2020,Puryear2020,Aiemjoy2020,Bell-Gorrod2020,Amusa2021,Reilly2021,Kang2021,Wang2021,Torres2021a,Mozafar2021,Mehta2021,Nardone2021,Lee2021,Rossides2021,Shiba2021,Zou2021,Goel2021,Kahkoska2021,Beydoun2021,Chen2022,Crowner2022,Isfordink2022,Ikeda2022a,Ikeda2022b} causation,\cite{Rosenblum2009,Padula2012,Leslie2014,Schnitzer2014,Brown2015,Gianfrancesco2016,Tran2016,Herrera2017,Skeem2017,Skeem2018ComparingIllness,Izano2019,Lim2019,Rodriguez2019,Gianfrancesco2019,Papadopoulou2019,Veit2020,Kagawa2020,Torres2020,Westling2020,Decruyenaere2020,Clare2020,Abdollahpour2021,Torres2021b,Almasi2021,Kerschberger2021,Dadi2021,Sun2021,Chavda2022,Akhtar2022,Wong2022,MorenoBetancur2022} and variable importance. \cite{Bembom2009,Legrand2013,Hsu2016,Sun2021} It has been used to analyse data with varying numbers of observations, from less than 100 to over hundreds of thousands, from clinical trials, cohort studies,\cite{Rosenblum2009,Mackey2011,Schnitzer2014,Brown2015,Gianfrancesco2016,Skeem2017,Sukul2017,Torres2018,Izano2019,Lim2019,Gianfrancesco2019,Casey2019,Torres2019,Veit2020,Ehrlich2020,Kempker2020,Torres2020,Westling2020,Clare2020,Bell-Gorrod2020,Reilly2021,Wang2021,Torres2021b,Kerschberger2021,Lee2021,Shiba2021,Dadi2021,Zou2021,Sun2021,Goel2021,Kahkoska2021,Chavda2022,Chen2022,Crowner2022,Isfordink2022,Ikeda2022a,Ikeda2022b,MorenoBetancur2022} and observational studies.

\newpage
    \newpage

\section{Discussion}\label{sec4}

    We aimed to investigate the use of the TMLE framework in epidemiology and public health research and to describe the uptake of its methodological developments since its inception in 2006. We focused on TMLEs for point treatment, time-to-event/survival, and longitudinal exposure-outcome relationships. We did not discuss TMLE methodological developments and software for mediation analyses.\cite{Zheng2012,Lendle2013b,Rudolph2018} We found that the TMLE framework and its different estimators were implemented in at least 81 epidemiological observational studies. The majority of these studies have come from the US, many of which are from the University of California, Berkeley. Recently, the use of TMLE has spread across the world. Until 2016, TMLE in observational studies was used by select groups of expertise and disciplines, such as biostatisticians or epidemiologists in academia exploring noncommunicable and infectious diseases, or behavioural epidemiology. From 2016 onward, there was a faster uptake amongst a wider range of expertise and disciplines. There is potential for even wider dissemination and acceptance, both geographically and in some specific disease areas or epidemiological disciplines. We hope this review of explicit and applied examples will contribute to enhancing the relevance of the TMLE framework and increasing its uptake and acceptance in settings where challenges with regard to data, unrealistic assumptions, or subject-matter knowledge lend themselves to the framework. \\
    
    Initially, causal inference methods and estimators relied on parametric modelling assumptions but, to quote Box (1976), ``all models are wrong but some are useful".\cite{Box1976ScienceStatistics} It highlights that model misspecification was and remains a challenge, even with ever-growing datasets and computing power: exclusion of the unknown data-generating distribution leads to biased results. Semi-parametric and non-parametric causal inference estimators, such as AIPTW, double-debiased,\cite{Chernozhukov2018Double/debiasedParameters} and TMLE\cite{vanderLaan2006} aim to provide the least biased estimate of the effect of an exposure on an outcome.\cite{Zivich2021,Smith2022} Maximum Likelihood Estimation (MLE) based methods (stratification, propensity score and parametric regression adjustment) and other estimating equations (AIPTW) do not have all of the properties of TMLE, and evidence shows that they under-perform in comparison to TMLE in specific settings.\cite{vanderLaan2006,Rose2011,Luque-Fernandez2018,Smith2022} For more detailed comparisons between the different methods, the interested reader is referred to Chapter 6 of van der Laan and Rose’s TMLE textbook.\cite{vanderLaan2011TargetedLearning} It is important to highlight that in contrast to the AIPTW estimator, TMLE respects the global constraints of the statistical model (i.e., $P(0 < Y < 1) = 1$) and solves the score estimation equations being equal to zero using substitution and fluctuation approaches.\cite{Gruber2010b} TMLE augments the initial estimates to obtain an optimal bias-variance trade-off for the target estimand of interest and produces a well-defined, unbiased, efficient substitution estimator. Furthermore, the targeting step (i.e., update of the initial estimate) may remove finite sample bias. Lastly, the TMLE framework can be tailored to specific research questions that are difficult to answer using other causal inference methods, such as rare diseases,\cite{Balzer2016a, Benkeser2018} ordinal\cite{Diaz2016} or continuous exposures,\cite{Diaz2021NonparametricPolicies} dynamic treatment regimes,\cite{Peterson2014} and missing outcome data.\cite{Diaz2017} \\
    
    We argue that dissemination of any new statistical methodology relies on five key factors: (i) software availability, (ii) accessibility of available material (e.g., quality of software help files, language used in publications, etc.), (iii) number of experts in the area, (iv) teaching, and (v) collaborations. In the following, we discuss the dissemination of TMLE with regard to each of them. \\
    
    \noindent \paragraph{(i) Software availability:} \, \\
    Various TMLEs have been developed for complex study designs, such as those with time-to-event outcomes, case-control studies, hierarchical data structures (including cluster randomised trials), longitudinal data, and time-dependent confounding. These methodological developments were accompanied by the release of R software packages, increasing the usability of TMLE. Such software developments include the \textit{SuperLearner}\cite{VanDerLaan2007SuperLearner} R package in 2007 and the \textit{tmle} R package in 2012.\cite{grubertmle} TMLE software for survival analysis (\textit{survtmle}),\cite{benkeser2017improved,benkeser2017survtmle} longitudinal data (\textit{ltmle}),\cite{Lendle2017Ltmle:Data} doubly-robust confidence intervals (\textit{drtmle}),\cite{drtmlepackage} and estimation of the survival curve under static, dynamic and stochastic interventions (\textit{stremr})\cite{sofrygin2017stremr} were implemented in 2017. To match the expanding framework, further software developments occurred in the following years, such as the \textit{tlverse} suite of software packages for Targeted Learning (\url{https://tlverse.org/tlverse-handbook/}), which includes R packages for cross-validation (\textit{origami}),\cite{coyle2018origami} highly adaptive lasso (HAL, \textit{hal9001}),\cite{coyle2022hal9001-rpkg,hejazi2020hal9001-joss} super learning (\textit{sl3}),\cite{coyle2021sl3-rpkg} and TMLEs for a range of target estimands, such as effects under static interventions on an exposure (\textit{tmle3}),\cite{coyle2021tmle3-rpkg} optimal dynamic treatment regimes for binary and categorical exposures (\textit{tmle3mopttx}),\cite{malenica2022tmle3mopttx} and stochastic treatment regimes that shift the treatment mechanism of a continuous exposure (\textit{tmle3shift}).\cite{hejazi2021tmle3shift-rpkg} Additional recently developed packages in R include \textit{ctmle} for collaborative TMLE,\cite{ctmle2017} \textit{haldensify} for conditional density estimation with HAL, \cite{Hejazi2022Haldensify:R,Ertefaie2020NonparametricLasso} \textit{txshift} for estimating causal effects of stochastic interventions,\cite{hejazi2020efficient,hejazi2020txshift-joss,hejazi2022txshift-rpkg} and \textit{lmtp} for longitudinal modified treatment policies.\cite{Diaz2021NonparametricPolicies}\\
    
    \noindent Although the TMLE framework is well developed in the R software, applied epidemiological research is performed in several other software languages, such as Stata, Python, and SAS. TMLE implementations for binary point exposure and outcome studies are available in all of these languages. A SAS macro for the general implementation of TMLE was programmed in 2016.\cite{Pang2016a} TMLE has been developed for the Python software language in the library \textit{zEpid}.\citep{zepid} The number of applied researchers in epidemiological studies using Python is relatively low but is increasing; thus, this tool is not currently widely used among applied health sciences researchers. Further development could improve on software packages in the widely used statistical software in health sciences and econometrics, such as Stata.\cite{Luque-Fernandez2021ELTMLE:Estimation} Nonetheless, the development version of the user-written Stata command \textit{eltmle} is currently available to Stata users.\cite{Luque-Fernandez2021ELTMLE:Estimation} Not all features of TMLE are available in this Stata command, such as longitudinal analysis and cross-validated TMLE. Additionally, \textit{eltmle} provides ensemble learning capabilities by accessing the \textit{SuperLearner} R package. 
    Lastly, any new software development needs to have a friendly user interface, together with standard programming features to be easily disseminated and quickly adopted. \\
    
    \noindent \paragraph{(ii) Accessibility of available material:} \, \\
    The TMLE framework is a series of potentially statistically-complex modelling approaches and computational algorithms, grounded in statistical theory that requires a solid understanding of highly advanced statistics (i.e., theory for semi-parametric estimation, asymptotics, efficiency, empirical processes, functional analyses, and statistical inference). Tutorials in a more lay language targeting applied researchers and epidemiologists have become more common over the past five years and the uptake of TMLE is expected to increase in the future because of them.\cite{Gruber2015a,Pang2016a,Pearl2016,Schuler2017,Kreif2017,Luque-Fernandez2018,Almasi2018,Smith2022,Rose2011,Phillips2022PracticalLearner} Their beneficial impact is evident from this review, as these articles are highly referenced in applied work, from the year of their publication, alongside more methodologically heavy contributions to start with, and as sole references in later years. This shows evidence of the importance of speaking the language of the target audience and disseminating advanced mathematical statistics and algorithms from an applied perspective. \\
    
    \noindent Additionally, the gradual dissemination of the TMLE framework was evident from our systematic review of the methods sections of the 81 selected manuscripts. We observed that papers published in the early years lay out their TMLE strategy and carefully describe each step in the methods section; whereas, more recently, publications of applied research have placed details of the methods in appendices (or supplementary material) and only cite tutorials and software packages. This shows that the community (e.g., authors, editors, reviewers, readers, etc.) is now aware of the TMLE framework, its utility, and its advantages. A wide range of journals have published the applied research articles studied here, from non-specific public health journals to statistical or disease-specific journals.\\

    \noindent \paragraph{(iii) Experts:} \, \\
    Dissemination outside the US needs further work, as evidenced in our systematic review. We have shown that the TMLE framework appears to be well consolidated in the US, and adoption from Europe and other regions are lower in comparison. This may be related to the delayed introduction of causal inference education outside the US. Fostering targeted local seminars and dedicated short courses for the interested applied audience could be a useful strategy to disseminate the framework. Disease- or discipline-specific experts would be useful for the wider distribution of the methods in specific areas that would benefit from improved methodology. \\
    
    \noindent TMLE remains dominant in non-communicable or infectious disease epidemiology compared to other disciplines, but it has high applicability in many disciplines and its use has increased in several of them. The slower uptake of the TMLE framework in other disciplines might be due to a lack of empirical examples of how one performed and interpreted statistical analyses using TMLE in a specific disease area. We aimed to provide such a showcase of the application of the methods in specific settings, based on the available literature, and we demonstrated how the framework was successfully used to advance research by providing robust results. We believe interested readers will find it useful to refer to the studies that faced similar challenges, or were based in settings comparable to theirs. \\
    
    \noindent \paragraph{(iv) Teaching:} \, \\
    There have been tremendous efforts of dissemination of causal inference methods across disciplines, with a particular emphasis on epidemiology and econometrics sciences in the US during the last 20 years. Most graduate programs in epidemiology have included the teaching of causal inference as a leading topic in the field. In Europe, the trends have not been as fast-paced and for a long time, introductions to causal inference methods have mainly been provided through week-long intensive short courses and at international conferences. These different approaches have major impacts on how quickly the methods are adopted by the community of researchers, journal editors, public health groups, and regulatory agencies. In recent years, there has been a development and acceptance of real-world evidence in various public-health fields, such as the Food and Drug Administration's 21st Century Cures Act of 2016 in the US, which specifically promotes the use of causal inference methodology and designs, such as the emulated trial and TMLE frameworks.\cite{GruberArxiv,SentinelInnov,gruber2022developing}\\ 
    
    \noindent \paragraph{(v) Collaborations:} \, \\
    The Center for Targeted Machine Learning and Causal Inference (CTML) is an interdisciplinary research centre at the University of California at Berkeley that is focused on applications of causal inference and targeted learning. The CTML mission is to advance, implement and disseminate methodology to address problems arising in public health and clinical medicine (\url{https://ctml.berkeley.edu/home}). CTML provides a great resource for courses, ongoing research, partners, collaborators, and Berkeley faculty members involved in TMLE. CTML sponsors include the Danish multinational pharmaceutical company, Novo Nordisk A/S, the Patient-Centered Outcomes Research Institute (pcori), Kaiser Permanente, the US National Institutes of Health, and the Bill and Melinda Gates Foundation. Academic partners include the University of Washington, University of Copenhagen, UCLA David Geffen School of Medicine, University of California at San Francisco, and Monash University.\\ 


    \paragraph{Conclusions}
    Evidence shows that cross-validated, double-robust, efficient and unbiased estimators are at the forefront of causal inference and statistics, as they aim to avoid model misspecification, bias and invalid inference. The TMLE framework for causal and statistical inference was first developed in 2006 and its expansion in applied studies arose in 2018 via applied epidemiological work, tutorials and user-friendly software. The theoretical properties and practical benefits of the TMLE framework have been highlighted across different fields of applied research (such as various epidemiological, public health and clinical disciplines). More can be done to reach a wider audience across varied disease areas and scientific fields (e.g., genomics, econometrics, political and sociological sciences), including the development of software packages outside the R software, tutorial articles as well as seminars and courses targeted to audiences in specific disciplines, lay-language demonstration, such as by example, of the benefits of TMLE in improving epidemiological output, to name only a few ideas. Many recent TMLE developments answer a variety of methodological problems that expand across scientific disciplines and further efforts can be made to disseminate the framework. This would facilitate the conscientious application of TMLE for causal inference and statistical data analyses, so more researchers could use it in their applied work to minimise the risk of reporting misleading results that are biased due to misspecification.

\newpage
\section*{Funding}
This work was supported by the Medical Research Council [grant number MR/W021021/1]. A CC BY or equivalent licence is applied to the Author Accepted Manuscript (AAM) arising from this submission, in accordance with the grant’s open access conditions. Camille Maringe is supported by a Cancer Research UK Population Research Committee Programme Award (C7923/A29018).

\section*{Authors contributions}
The article arose from the motivation to disseminate the principles of modern epidemiology among clinicians and applied researchers. All authors developed the concept and wrote the first draft of the article. MJS and CM reviewed the literature. MJS, RVP, MALF and CM drafted and revised the manuscript. RVP, SG and MJL provided comments on the draft manuscript. RVP contributed to drafting some sections. All authors read and approved the final version of the manuscript. CM is the guarantor of the article.

\section*{Acknowledgments}
The motivation and some parts of the manuscript come from MALF's work in a visiting academic position in the Division of Biostatistics at the Berkeley School of Public Health in 2019.
We acknowledge Dr Susan Gruber for her careful read over one draft manuscript and Professor Mark J. van der Laan for his additional guidance.\\

For the purpose of open access, the authors have applied a Creative Commons Attribution (CC BY) licence to any Author Accepted Manuscript version arising.

\newpage

\bibliography{wileyNJD-AMA}

\begin{thebibliography}{100}
\providecommand \doibase [0]{http://dx.doi.org/}%

\bibitem{Smith2022}
Smith MJ, Mansournia MA, Maringe C, et al. Introduction to computational causal
  inference using reproducible Stata, R, and Python code: A tutorial. {\it Stat
  Med} 2022\string; 41(2)\string: 407-432.
\newblock \href {\doibase 10.1002/sim.9234} {doi: 10.1002/sim.9234}

\bibitem{vanderLaan2011TargetedLearning}
\text{van der Laan} MJ, Rose S. {Targeted Learning}.  2011.
\newblock \href {\doibase 10.1007/978-1-4419-9782-1} {doi:
  10.1007/978-1-4419-9782-1}

\bibitem{vanderLaan2006}
\text{van der Laan} MJ, Rubin D. Targeted maximum likelihood learning. {\it The
  International Journal of Biostatistics} 2006\string; 2(1).
\newblock \href {\doibase https://doi.org/10.2202/1557-4679.1043} {doi:
  https://doi.org/10.2202/1557-4679.1043}

\bibitem{VanDerLaan2007SuperLearner}
\text{van der Laan} MJ, Polley EC, Hubbard AE. {Super learner}. {\it
  Statistical applications in genetics and molecular biology} 2007\string;
  6(1).
\newblock \href {\doibase 10.2202/1544-6115.1309} {doi: 10.2202/1544-6115.1309}

\bibitem{Luque-Fernandez2018}
Luque-Fernandez MA, Schomaker M, Rachet B, Schnitzer ME. Targeted maximum
  likelihood estimation for a binary treatment: A tutorial. {\it Stat Med}
  2018\string; 37(16)\string: 2530-2546.
\newblock \href {\doibase 10.1002/sim.7628} {doi: 10.1002/sim.7628}

\bibitem{Gruber2015a}
Gruber S. Targeted Learning in Healthcare Research. {\it Big Data} 2015\string;
  3(4)\string: 211-8.
\newblock \href {\doibase 10.1089/big.2015.0025} {doi: 10.1089/big.2015.0025}

\bibitem{Schuler2017}
Schuler MS, Rose S. Targeted Maximum Likelihood Estimation for Causal Inference
  in Observational Studies. {\it Am J Epidemiol} 2017\string; 185(1)\string:
  65-73.
\newblock \href {\doibase 10.1093/aje/kww165} {doi: 10.1093/aje/kww165}

\bibitem{vanderLaan2018TargetedLearning}
\text{van der Laan} MJ, Rose S. {Targeted learning in data science: causal
  inference for complex longitudinal studies}.  2011.
\newblock \href {\doibase 10.1007/978-3-319-65304-4} {doi:
  10.1007/978-3-319-65304-4}

\bibitem{UCBerkeley2022American2022}
{UC Berkeley} . {American Causal Inference Conference (2022)}.
  https://ctml.berkeley.edu/american-causal-inference-conference-2022;  2022

\bibitem{UCBerkeley2022AmericanResults}
{UC Berkeley} . {American Causal Inference Conference (2022): Results}.
  https://acic2022.mathematica.org/results;  2022

\bibitem{Luque-Fernandez2018Data-AdaptivePresentation}
Luque-Fernandez MA, Belot A, Valeri L, Cerulli G, Maringe C, Rachet B.
  {Data-Adaptive Estimation for Double-Robust Methods in Population-Based
  Cancer Epidemiology: Risk Differences for Lung Cancer Mortality by Emergency
  Presentation}. {\it American Journal of Epidemiology} 2018\string;
  187(4)\string: 871--878.
\newblock \href {\doibase 10.1093/AJE/KWX317} {doi: 10.1093/AJE/KWX317}

\bibitem{Bembom2009}
Bembom O, Petersen ML, Rhee SY, et al. Biomarker discovery using targeted
  maximum-likelihood estimation: application to the treatment of
  antiretroviral-resistant HIV infection. {\it Stat Med} 2009\string;
  28(1)\string: 152-72.
\newblock \href {\doibase 10.1002/sim.3414} {doi: 10.1002/sim.3414}

\bibitem{Rosenblum2009}
Rosenblum M, Deeks SG, \text{van der Laan} MJ, Bangsberg DR. The risk of
  virologic failure decreases with duration of HIV suppression, at greater than
  50\% adherence to antiretroviral therapy. {\it PLoS One} 2009\string;
  4(9)\string: e7196.
\newblock \href {\doibase 10.1371/journal.pone.0007196} {doi:
  10.1371/journal.pone.0007196}

\bibitem{Mackey2011}
Mackey DC, Hubbard AE, Cawthon PM, Cauley JA, Cummings SR, Tager IB. Usual
  physical activity and hip fracture in older men: an application of
  semiparametric methods to observational data. {\it Am J Epidemiol}
  2011\string; 173(5)\string: 578-86.
\newblock \href {\doibase 10.1093/aje/kwq405} {doi: 10.1093/aje/kwq405}

\bibitem{Padula2012}
Padula AM, Mortimer K, Hubbard A, Lurmann F, Jerrett M, Tager IB. Exposure to
  traffic-related air pollution during pregnancy and term low birth weight:
  estimation of causal associations in a semiparametric model. {\it Am J
  Epidemiol} 2012\string; 176(9)\string: 815-24.
\newblock \href {\doibase 10.1093/aje/kws148} {doi: 10.1093/aje/kws148}

\bibitem{Leslie2014}
Leslie HH, Karasek DA, Harris LF, et al. Cervical cancer precursors and
  hormonal contraceptive use in HIV-positive women: application of a causal
  model and semi-parametric estimation methods. {\it PLoS One} 2014\string;
  9(6)\string: e101090.
\newblock \href {\doibase 10.1371/journal.pone.0101090} {doi:
  10.1371/journal.pone.0101090}

\bibitem{Schnitzer2014}
Schnitzer ME, Moodie EE, \text{van der Laan} MJ, Platt RW, Klein MB. Modeling
  the impact of hepatitis C viral clearance on end-stage liver disease in an
  HIV co-infected cohort with targeted maximum likelihood estimation. {\it
  Biometrics} 2014\string; 70(1)\string: 144-52.
\newblock \href {\doibase 10.1111/biom.12105} {doi: 10.1111/biom.12105}

\bibitem{Brown2015}
Brown DM, Petersen M, Costello S, et al. Occupational Exposure to PM2.5 and
  Incidence of Ischemic Heart Disease: Longitudinal Targeted Minimum Loss-based
  Estimation. {\it Epidemiology} 2015\string; 26(6)\string: 806-14.
\newblock \href {\doibase 10.1097/ede.0000000000000329} {doi:
  10.1097/ede.0000000000000329}

\bibitem{Gianfrancesco2016}
Gianfrancesco MA, Balzer L, Taylor KE, et al. Genetic risk and longitudinal
  disease activity in systemic lupus erythematosus using targeted maximum
  likelihood estimation. {\it Genes Immun} 2016\string; 17(6)\string: 358-62.
\newblock \href {\doibase 10.1038/gene.2016.33} {doi: 10.1038/gene.2016.33}

\bibitem{Hsu2016}
Hsu LI, Briggs F, Shao X, et al. Pathway Analysis of Genome-wide Association
  Study in Childhood Leukemia among Hispanics. {\it Cancer Epidemiol Biomarkers
  Prev} 2016\string; 25(5)\string: 815-22.
\newblock \href {\doibase 10.1158/1055-9965.Epi-15-0528} {doi:
  10.1158/1055-9965.Epi-15-0528}

\bibitem{Ahern2016}
Ahern J, Karasek D, Luedtke AR, Bruckner TA, \text{van der Laan} MJ.
  Racial/Ethnic Differences in the Role of Childhood Adversities for Mental
  Disorders Among a Nationally Representative Sample of Adolescents. {\it
  Epidemiology} 2016\string; 27(5)\string: 697-704.
\newblock \href {\doibase 10.1097/ede.0000000000000507} {doi:
  10.1097/ede.0000000000000507}

\bibitem{Salihu2016}
Salihu HM, Das R, Morton L, et al. Racial Differences in DNA-Methylation of CpG
  Sites Within Preterm-Promoting Genes and Gene Variants. {\it Matern Child
  Health J} 2016\string; 20(8)\string: 1680-7.
\newblock \href {\doibase 10.1007/s10995-016-1967-3} {doi:
  10.1007/s10995-016-1967-3}

\bibitem{Tran2016}
Tran L, Yiannoutsos CT, Musick BS, et al. Evaluating the Impact of a HIV
  Low-Risk Express Care Task-Shifting Program: A Case Study of the Targeted
  Learning Roadmap. {\it Epidemiol Methods} 2016\string; 5(1)\string: 69-91.
\newblock \href {\doibase 10.1515/em-2016-0004} {doi: 10.1515/em-2016-0004}

\bibitem{Davis2017}
Davis FM, Sutzko DC, Grey SF, et al. Predictors of surgical site infection
  after open lower extremity revascularization. {\it J Vasc Surg} 2017\string;
  65(6)\string: 1769-1778.e3.
\newblock \href {\doibase 10.1016/j.jvs.2016.11.053} {doi:
  10.1016/j.jvs.2016.11.053}

\bibitem{Skeem2017}
Skeem JL, Manchak S, Montoya L. Comparing Public Safety Outcomes for
  Traditional Probation vs Specialty Mental Health Probation. {\it JAMA
  Psychiatry} 2017\string; 74(9)\string: 942-948.
\newblock \href {\doibase 10.1001/jamapsychiatry.2017.1384} {doi:
  10.1001/jamapsychiatry.2017.1384}

\bibitem{Sukul2017}
Sukul D, Seth M, Dixon SR, Khandelwal A, LaLonde TA, Gurm HS. Contemporary
  Trends and Outcomes Associated With the Preprocedural Use of Oral P2Y12
  Inhibitors in Patients Undergoing Percutaneous Coronary Intervention:
  Insights From the Blue Cross Blue Shield of Michigan Cardiovascular
  Consortium (BMC2). {\it J Invasive Cardiol} 2017\string; 29(10)\string:
  340-351.

\bibitem{Akosile2018ReassessingEstimation}
Akosile M, Zhu H, Zhang S, Johnson NP, Lai D, Zhu H. {Reassessing the
  Effectiveness of Right Heart Catheterization (RHC) in the Initial Care of
  Critically Ill Patients using Targeted Maximum Likelihood Estimation}. {\it
  International Journal of Clinical Biostatistics and Biometrics} 2018\string;
  4(1).
\newblock \href {\doibase 10.23937/2469-5831/1510018} {doi:
  10.23937/2469-5831/1510018}

\bibitem{Platt2018}
Platt JM, McLaughlin KA, Luedtke AR, Ahern J, Kaufman AS, Keyes KM. Targeted
  Estimation of the Relationship Between Childhood Adversity and Fluid
  Intelligence in a US Population Sample of Adolescents. {\it Am J Epidemiol}
  2018\string; 187(7)\string: 1456-1466.
\newblock \href {\doibase 10.1093/aje/kwy006} {doi: 10.1093/aje/kwy006}

\bibitem{Pearl2018}
Pearl M, Ahern J, Hubbard A, et al. Life-course neighbourhood opportunity and
  racial-ethnic disparities in risk of preterm birth. {\it Paediatr Perinat
  Epidemiol} 2018\string; 32(5)\string: 412-419.
\newblock \href {\doibase 10.1111/ppe.12482} {doi: 10.1111/ppe.12482}

\bibitem{Skeem2018}
Skeem JL, Montoya L, Manchak SM. Comparing Costs of Traditional and Specialty
  Probation for People With Serious Mental Illness. {\it Psychiatr Serv}
  2018\string; 69(8)\string: 896-902.
\newblock \href {\doibase 10.1176/appi.ps.201700498} {doi:
  10.1176/appi.ps.201700498}

\bibitem{Mosconi2018}
Mosconi L, Rahman A, Diaz I, et al. Increased Alzheimer's risk during the
  menopause transition: A 3-year longitudinal brain imaging study. {\it PLoS
  One} 2018\string; 13(12)\string: e0207885.
\newblock \href {\doibase 10.1371/journal.pone.0207885} {doi:
  10.1371/journal.pone.0207885}

\bibitem{Torres2018}
Torres JM, Rudolph KE, Sofrygin O, Glymour MM, Wong R. Longitudinal
  associations between having an adult child migrant and depressive symptoms
  among older adults in the Mexican Health and Aging Study. {\it Int J
  Epidemiol} 2018\string; 47(5)\string: 1432-1442.
\newblock \href {\doibase 10.1093/ije/dyy112} {doi: 10.1093/ije/dyy112}

\bibitem{Izano2019}
Izano MA, Sofrygin OA, Picciotto S, Bradshaw PT, Eisen EA. Metalworking Fluids
  and Colon Cancer Risk: Longitudinal Targeted Minimum Loss-based Estimation.
  {\it Environ Epidemiol} 2019\string; 3(1)\string: e035.
\newblock \href {\doibase 10.1097/ee9.0000000000000035} {doi:
  10.1097/ee9.0000000000000035}

\bibitem{Yu2019}
Yu YH, Bodnar LM, Brooks MM, Himes KP, Naimi AI. Comparison of Parametric and
  Nonparametric Estimators for the Association Between Incident Prepregnancy
  Obesity and Stillbirth in a Population-Based Cohort Study. {\it Am J
  Epidemiol} 2019\string; 188(7)\string: 1328-1336.
\newblock \href {\doibase 10.1093/aje/kwz081} {doi: 10.1093/aje/kwz081}

\bibitem{Rudolph2019}
Rudolph KE, Shev A, Paksarian D, et al. Environmental noise and sleep and
  mental health outcomes in a nationally representative sample of urban US
  adolescents. {\it Environ Epidemiol} 2019\string; 3(4)\string: e056.
\newblock \href {\doibase 10.1097/ee9.0000000000000056} {doi:
  10.1097/ee9.0000000000000056}

\bibitem{Lim2019}
Lim S, Tellez M, Ismail AI. Estimating a Dynamic Effect of Soda Intake on
  Pediatric Dental Caries Using Targeted Maximum Likelihood Estimation Method.
  {\it Caries Res} 2019\string; 53(5)\string: 532-540.
\newblock \href {\doibase 10.1159/000497359} {doi: 10.1159/000497359}

\bibitem{Berkowitz2019}
Berkowitz SA, Basu S, Gundersen C, Seligman HK. State-Level and County-Level
  Estimates of Health Care Costs Associated with Food Insecurity. {\it Prev
  Chronic Dis} 2019\string; 16\string: E90.
\newblock \href {\doibase 10.5888/pcd16.180549} {doi: 10.5888/pcd16.180549}

\bibitem{Gianfrancesco2019}
Gianfrancesco MA, Trupin L, Shiboski S, et al. Smoking Is Associated with
  Higher Disease Activity in Rheumatoid Arthritis: A Longitudinal Study
  Controlling for Time-varying Covariates. {\it J Rheumatol} 2019\string;
  46(4)\string: 370-375.
\newblock \href {\doibase 10.3899/jrheum.180262} {doi: 10.3899/jrheum.180262}

\bibitem{Casey2019}
Casey JA, Goin DE, Rudolph KE, et al. Unconventional natural gas development
  and adverse birth outcomes in Pennsylvania: The potential mediating role of
  antenatal anxiety and depression. {\it Environ Res} 2019\string; 177\string:
  108598.
\newblock \href {\doibase 10.1016/j.envres.2019.108598} {doi:
  10.1016/j.envres.2019.108598}

\bibitem{Torres2019}
Torres JM, Rudolph KE, Sofrygin O, Wong R, Walter LC, Glymour MM. Having an
  Adult Child in the United States, Physical Functioning, and Unmet Needs for
  Care Among Older Mexican Adults. {\it Epidemiology} 2019\string;
  30(4)\string: 553-560.
\newblock \href {\doibase 10.1097/ede.0000000000001016} {doi:
  10.1097/ede.0000000000001016}

\bibitem{You2019ApplicationMexico}
You Y, Doubova SV, Pinto-Masis D, P{\'{e}}rez-Cuevas R, Borja-Aburto VH,
  Hubbard A. {Application of machine learning methodology to assess the
  performance of DIABETIMSS program for patients with type 2 diabetes in family
  medicine clinics in Mexico}. {\it BMC Medical Informatics and Decision
  Making} 2019\string; 19(1).
\newblock \href {\doibase 10.1186/S12911-019-0950-5} {doi:
  10.1186/S12911-019-0950-5}

\bibitem{Izano2020}
Izano MA, Cushing LJ, Lin J, et al. The association of maternal psychosocial
  stress with newborn telomere length. {\it PLoS One} 2020\string;
  15(12)\string: e0242064.
\newblock \href {\doibase 10.1371/journal.pone.0242064} {doi:
  10.1371/journal.pone.0242064}

\bibitem{Ehrlich2020}
Ehrlich SF, Neugebauer RS, Feng J, Hedderson MM, Ferrara A. Exercise During the
  First Trimester and Infant Size at Birth: Targeted Maximum Likelihood
  Estimation of the Causal Risk Difference. {\it Am J Epidemiol} 2020\string;
  189(2)\string: 133-145.
\newblock \href {\doibase 10.1093/aje/kwz213} {doi: 10.1093/aje/kwz213}

\bibitem{Bodnar2020}
Bodnar LM, Cartus AR, Kirkpatrick SI, et al. Machine learning as a strategy to
  account for dietary synergy: an illustration based on dietary intake and
  adverse pregnancy outcomes. {\it Am J Clin Nutr} 2020\string; 111(6)\string:
  1235-1243.
\newblock \href {\doibase 10.1093/ajcn/nqaa027} {doi: 10.1093/ajcn/nqaa027}

\bibitem{Kagawa2020}
Kagawa RMC, Pear VA, Rudolph KE, Keyes KM, Cerdá M, Wintemute GJ. Distress
  level and daily functioning problems attributed to firearm victimization:
  sociodemographic-specific responses. {\it Ann Epidemiol} 2020\string;
  41\string: 35-42.e3.
\newblock \href {\doibase 10.1016/j.annepidem.2019.12.002} {doi:
  10.1016/j.annepidem.2019.12.002}

\bibitem{Kempker2020}
Kempker RR, Mikiashvili L, Zhao Y, et al. Clinical Outcomes Among Patients With
  Drug-resistant Tuberculosis Receiving Bedaquiline- or Delamanid-Containing
  Regimens. {\it Clin Infect Dis} 2020\string; 71(9)\string: 2336-2344.
\newblock \href {\doibase 10.1093/cid/ciz1107} {doi: 10.1093/cid/ciz1107}

\bibitem{Yu2020}
Yu YH, Bodnar LM, Himes KP, Brooks MM, Naimi AI. Association of Overweight and
  Obesity Development Between Pregnancies With Stillbirth and Infant Mortality
  in a Cohort of Multiparous Women. {\it Obstet Gynecol} 2020\string;
  135(3)\string: 634-643.
\newblock \href {\doibase 10.1097/aog.0000000000003677} {doi:
  10.1097/aog.0000000000003677}

\bibitem{Puryear2020}
Puryear SB, Balzer LB, Ayieko J, et al. Associations between alcohol use and
  HIV care cascade outcomes among adults undergoing population-based HIV
  testing in East Africa. {\it Aids} 2020\string; 34(3)\string: 405-413.
\newblock \href {\doibase 10.1097/qad.0000000000002427} {doi:
  10.1097/qad.0000000000002427}

\bibitem{Torres2020}
Torres JM, Sofrygin O, Rudolph KE, Haan MN, Wong R, Glymour MM. US Migration
  Status of Adult Children and Cognitive Decline Among Older Parents Who Remain
  in Mexico. {\it Am J Epidemiol} 2020\string; 189(8)\string: 761-769.
\newblock \href {\doibase 10.1093/aje/kwz277} {doi: 10.1093/aje/kwz277}

\bibitem{Westling2020}
Westling T, Cowden C, Mwananyanda L, et al. Impact of chlorhexidine baths on
  suspected sepsis and bloodstream infections in hospitalized neonates in
  Zambia. {\it Int J Infect Dis} 2020\string; 96\string: 54-60.
\newblock \href {\doibase 10.1016/j.ijid.2020.03.043} {doi:
  10.1016/j.ijid.2020.03.043}

\bibitem{Aiemjoy2020}
Aiemjoy K, Aragie S, Wittberg DM, et al. Seroprevalence of antibodies against
  Chlamydia trachomatis and enteropathogens and distance to the nearest water
  source among young children in the Amhara Region of Ethiopia. {\it PLoS Negl
  Trop Dis} 2020\string; 14(9)\string: e0008647.
\newblock \href {\doibase 10.1371/journal.pntd.0008647} {doi:
  10.1371/journal.pntd.0008647}

\bibitem{Reilly2021}
Reilly ME, Conti MS, Day J, et al. Modified Lapidus vs Scarf Osteotomy Outcomes
  for Treatment of Hallux Valgus Deformity. {\it Foot Ankle Int} 2021\string;
  42(11)\string: 1454-1462.
\newblock \href {\doibase 10.1177/10711007211013776} {doi:
  10.1177/10711007211013776}

\bibitem{Kang2021}
Kang L, Vij A, Hubbard A, Shaw D. The unintended impact of helmet use on
  bicyclists' risk-taking behaviors. {\it J Safety Res} 2021\string; 79\string:
  135-147.
\newblock \href {\doibase 10.1016/j.jsr.2021.08.014} {doi:
  10.1016/j.jsr.2021.08.014}

\bibitem{Torres2021a}
Torres JM, Mitchell UA, Sofrygin O, et al. Associations between spousal
  caregiving and health among older adults in Mexico: A targeted estimation
  approach. {\it Int J Geriatr Psychiatry} 2021\string; 36(5)\string: 775-783.
\newblock \href {\doibase 10.1002/gps.5477} {doi: 10.1002/gps.5477}

\bibitem{Torres2021b}
Torres LK, Hoffman KL, Oromendia C, et al. Attributable mortality of acute
  respiratory distress syndrome: a systematic review, meta-analysis and
  survival analysis using targeted minimum loss-based estimation. {\it Thorax}
  2021\string; 76(12)\string: 1176-1185.
\newblock \href {\doibase 10.1136/thoraxjnl-2020-215950} {doi:
  10.1136/thoraxjnl-2020-215950}

\bibitem{Mehta2021}
Mehta B, Brantner C, Williams N, et al. Primary Care Provider Density and
  Elective Total Joint Replacement Outcomes. {\it Arthroplast Today}
  2021\string; 10\string: 73-78.
\newblock \href {\doibase 10.1016/j.artd.2021.05.010} {doi:
  10.1016/j.artd.2021.05.010}

\bibitem{Nardone2021}
Nardone A, Rudolph KE, Morello-Frosch R, Casey JA. Redlines and Greenspace: The
  Relationship between Historical Redlining and 2010 Greenspace across the
  United States. {\it Environ Health Perspect} 2021\string; 129(1)\string:
  17006.
\newblock \href {\doibase 10.1289/ehp7495} {doi: 10.1289/ehp7495}

\bibitem{Lee2021}
Lee JO, Kapteyn A, Clomax A, Jin H. Estimating influences of unemployment and
  underemployment on mental health during the COVID-19 pandemic: who suffers
  the most?. {\it Public Health} 2021\string; 201\string: 48-54.
\newblock \href {\doibase 10.1016/j.puhe.2021.09.038} {doi:
  10.1016/j.puhe.2021.09.038}

\bibitem{Shiba2021}
Shiba K, Torres JM, Daoud A, et al. Estimating the Impact of Sustained Social
  Participation on Depressive Symptoms in Older Adults. {\it Epidemiology}
  2021\string; 32(6)\string: 886-895.
\newblock \href {\doibase 10.1097/ede.0000000000001395} {doi:
  10.1097/ede.0000000000001395}

\bibitem{Goel2021}
Goel AR, Bruce HA, Williams N, Alexiades G. Long-Term Effects of Hearing Aids
  on Hearing Ability in Patients with Sensorineural Hearing Loss. {\it J Am
  Acad Audiol} 2021\string; 32(6)\string: 374-378.
\newblock \href {\doibase 10.1055/s-0041-1731592} {doi: 10.1055/s-0041-1731592}

\bibitem{Kahkoska2021}
Kahkoska AR, Abrahamsen TJ, Alexander GC, et al. Association Between
  Glucagon-Like Peptide 1 Receptor Agonist and Sodium-Glucose Cotransporter 2
  Inhibitor Use and COVID-19 Outcomes. {\it Diabetes Care} 2021\string;
  44(7)\string: 1564-1572.
\newblock \href {\doibase 10.2337/dc21-0065} {doi: 10.2337/dc21-0065}

\bibitem{Beydoun2021}
Beydoun HA, Beydoun MA, Huang S, Eid SM, Zonderman AB. Hospitalization outcomes
  among brain metastasis patients receiving radiation therapy with or without
  stereotactic radiosurgery from the 2005-2014 Nationwide Inpatient Sample.
  {\it Sci Rep} 2021\string; 11(1)\string: 19209.
\newblock \href {\doibase 10.1038/s41598-021-98563-y} {doi:
  10.1038/s41598-021-98563-y}

\bibitem{Crowner2022}
Crowner JR, Marston WA, Freeman NLB, et al. The Society for Vascular Surgery
  Objective Performance Goals for Critical Limb Ischemia are attainable in
  select patients with ischemic wounds managed with wound care alone. {\it Ann
  Vasc Surg} 2022\string; 78\string: 28-35.
\newblock \href {\doibase 10.1016/j.avsg.2021.06.034} {doi:
  10.1016/j.avsg.2021.06.034}

\bibitem{Wong2022}
Wong AK, Balzer LB. State-Level Masking Mandates and COVID-19 Outcomes in the
  United States: A Demonstration of the Causal Roadmap. {\it Epidemiology}
  2022\string; 33(2)\string: 228-236.
\newblock \href {\doibase 10.1097/ede.0000000000001453} {doi:
  10.1097/ede.0000000000001453}

\bibitem{Legrand2013}
Legrand M, Pirracchio R, Rosa A, et al. Incidence, risk factors and prediction
  of post-operative acute kidney injury following cardiac surgery for active
  infective endocarditis: an observational study. {\it Crit Care} 2013\string;
  17(5)\string: R220.
\newblock \href {\doibase 10.1186/cc13041} {doi: 10.1186/cc13041}

\bibitem{Herrera2017}
Herrera R, Berger U, Ehrenstein vOS, et al. Estimating the Causal Impact of
  Proximity to Gold and Copper Mines on Respiratory Diseases in Chilean
  Children: An Application of Targeted Maximum Likelihood Estimation. {\it Int
  J Environ Res Public Health} 2017\string; 15(1).
\newblock \href {\doibase 10.3390/ijerph15010039} {doi: 10.3390/ijerph15010039}

\bibitem{Rodriguez2019}
Rodríguez-Molina D, Barth S, Herrera R, Rossmann C, Radon K, Karnowski V. An
  educational intervention to improve knowledge about prevention against
  occupational asthma and allergies using targeted maximum likelihood
  estimation. {\it Int Arch Occup Environ Health} 2019\string; 92(5)\string:
  629-638.
\newblock \href {\doibase 10.1007/s00420-018-1397-1} {doi:
  10.1007/s00420-018-1397-1}

\bibitem{Papadopoulou2019}
Papadopoulou E, Haug LS, Sakhi AK, et al. Diet as a Source of Exposure to
  Environmental Contaminants for Pregnant Women and Children from Six European
  Countries. {\it Environ Health Perspect} 2019\string; 127(10)\string: 107005.
\newblock \href {\doibase 10.1289/ehp5324} {doi: 10.1289/ehp5324}

\bibitem{Vauchel2019ImpactAnalysis}
Vauchel T, Pirracchio R, Chaussard M, et al. {Impact of an Acinetobacter
  baumannii outbreak on kidney events in a burn unit: A targeted machine
  learning analysis}. {\it American Journal of Infection Control} 2019\string;
  47(4)\string: 435--438.
\newblock \href {\doibase 10.1016/j.ajic.2018.09.010} {doi:
  10.1016/j.ajic.2018.09.010}

\bibitem{Veit2020}
Veit C, Herrera R, Weinmayr G, et al. Long-term effects of asthma medication on
  asthma symptoms: an application of the targeted maximum likelihood
  estimation. {\it BMC Med Res Methodol} 2020\string; 20(1)\string: 307.
\newblock \href {\doibase 10.1186/s12874-020-01175-9} {doi:
  10.1186/s12874-020-01175-9}

\bibitem{Decruyenaere2020}
Decruyenaere A, Steen J, Colpaert K, Benoit DD, Decruyenaere J, Vansteelandt S.
  The obesity paradox in critically ill patients: a causal learning approach to
  a casual finding. {\it Crit Care} 2020\string; 24(1)\string: 485.
\newblock \href {\doibase 10.1186/s13054-020-03199-5} {doi:
  10.1186/s13054-020-03199-5}

\bibitem{Rossides2021}
Rossides M, Kullberg S, Di~Giuseppe D, et al. Infection risk in sarcoidosis
  patients treated with methotrexate compared to azathioprine: A retrospective
  'target trial' emulated with Swedish real-world data. {\it Respirology}
  2021\string; 26(5)\string: 452-460.
\newblock \href {\doibase 10.1111/resp.14001} {doi: 10.1111/resp.14001}

\bibitem{Zou2021}
Zou R, El~Marroun H, Cecil C, et al. Maternal folate levels during pregnancy
  and offspring brain development in late childhood. {\it Clin Nutr}
  2021\string; 40(5)\string: 3391-3400.
\newblock \href {\doibase 10.1016/j.clnu.2020.11.025} {doi:
  10.1016/j.clnu.2020.11.025}

\bibitem{Isfordink2022}
Isfordink CJ, Smit C, Boyd A, et al. Low hepatitis C virus-viremia prevalence
  yet continued barriers to direct-acting antiviral treatment in people living
  with HIV in the Netherlands. {\it Aids} 2022\string; 36(6)\string: 773-783.
\newblock \href {\doibase 10.1097/qad.0000000000003159} {doi:
  10.1097/qad.0000000000003159}

\bibitem{Bell-Gorrod2020}
Bell-Gorrod H, Fox MP, Boulle A, et al. The Impact of Delayed Switch to
  Second-Line Antiretroviral Therapy on Mortality, Depending on Definition of
  Failure Time and CD4 Count at Failure. {\it Am J Epidemiol} 2020\string;
  189(8)\string: 811-819.
\newblock \href {\doibase 10.1093/aje/kwaa049} {doi: 10.1093/aje/kwaa049}

\bibitem{Amusa2021}
Amusa L, Zewotir T, North D, Kharsany ABM, Lewis L. Association of medical male
  circumcision and sexually transmitted infections in a population-based study
  using targeted maximum likelihood estimation. {\it BMC Public Health}
  2021\string; 21(1)\string: 1642.
\newblock \href {\doibase 10.1186/s12889-021-11705-9} {doi:
  10.1186/s12889-021-11705-9}

\bibitem{Kerschberger2021}
Kerschberger B, Boulle A, Kuwengwa R, Ciglenecki I, Schomaker M. The Impact of
  Same-Day Antiretroviral Therapy Initiation Under the World Health
  Organization Treat-All Policy. {\it Am J Epidemiol} 2021\string;
  190(8)\string: 1519-1532.
\newblock \href {\doibase 10.1093/aje/kwab032} {doi: 10.1093/aje/kwab032}

\bibitem{Dadi2021}
Dadi AF, Miller ER, Woodman RJ, Azale T, Mwanri L. Effect of perinatal
  depression on risk of adverse infant health outcomes in mother-infant dyads
  in Gondar town: a causal analysis. {\it BMC Pregnancy Childbirth}
  2021\string; 21(1)\string: 255.
\newblock \href {\doibase 10.1186/s12884-021-03733-5} {doi:
  10.1186/s12884-021-03733-5}

\bibitem{Mozafar2020}
Mozafar~Saadati H, Mehrabi Y, Sabour S, Mansournia MA, Hashemi~Nazari SS.
  Estimating the effects of body mass index and central obesity on stroke in
  diabetics and non-diabetics using targeted maximum likelihood estimation:
  Atherosclerosis Risk in Communities study. {\it Obes Sci Pract} 2020\string;
  6(6)\string: 628-637.
\newblock \href {\doibase 10.1002/osp4.447} {doi: 10.1002/osp4.447}

\bibitem{Abdollahpour2021}
Abdollahpour I, Nedjat S, Almasi-Hashiani A, Nazemipour M, Mansournia MA,
  Luque-Fernandez MA. Estimating the Marginal Causal Effect and Potential
  Impact of Waterpipe Smoking on Risk of Multiple Sclerosis Using the Targeted
  Maximum Likelihood Estimation Method: A Large, Population-Based Incident
  Case-Control Study. {\it Am J Epidemiol} 2021\string; 190(7)\string:
  1332-1340.
\newblock \href {\doibase 10.1093/aje/kwab036} {doi: 10.1093/aje/kwab036}

\bibitem{Mozafar2021}
Mozafar~Saadati H, Sabour S, Mansournia MA, Mehrabi Y, Hashemi~Nazari SS.
  Effect modification of general and central obesity by sex and age on
  cardiovascular outcomes: Targeted maximum likelihood estimation in the
  atherosclerosis risk in communities study. {\it Diabetes Metab Syndr}
  2021\string; 15(2)\string: 479-485.
\newblock \href {\doibase 10.1016/j.dsx.2021.02.024} {doi:
  10.1016/j.dsx.2021.02.024}

\bibitem{Almasi2021}
Almasi-Hashiani A, Nedjat S, Ghiasvand R, et al. The causal effect and impact
  of reproductive factors on breast cancer using super learner and targeted
  maximum likelihood estimation: a case-control study in Fars Province, Iran.
  {\it BMC Public Health} 2021\string; 21(1)\string: 1219.
\newblock \href {\doibase 10.1186/s12889-021-11307-5} {doi:
  10.1186/s12889-021-11307-5}

\bibitem{Akhtar2022}
Akhtar S, El-Muzaini H, Alroughani R. Recombinant hepatitis B vaccine uptake
  and multiple sclerosis risk: A marginal structural modeling approach. {\it
  Mult Scler Relat Disord} 2022\string; 58\string: 103487.
\newblock \href {\doibase 10.1016/j.msard.2022.103487} {doi:
  10.1016/j.msard.2022.103487}

\bibitem{Clare2020}
Clare PJ, Dobbins T, Bruno R, et al. The overall effect of parental supply of
  alcohol across adolescence on alcohol-related harms in early adulthood-a
  prospective cohort study. {\it Addiction} 2020\string; 115(10)\string:
  1833-1843.
\newblock \href {\doibase 10.1111/add.15005} {doi: 10.1111/add.15005}

\bibitem{Wang2021}
Wang L, Sun X, Jin C, Fan Y, Xue F. Identification of Tumor
  Microenvironment-Related Prognostic Biomarkers for Ovarian Serous Cancer
  3-Year Mortality Using Targeted Maximum Likelihood Estimation: A TCGA Data
  Mining Study. {\it Front Genet} 2021\string; 12\string: 625145.
\newblock \href {\doibase 10.3389/fgene.2021.625145} {doi:
  10.3389/fgene.2021.625145}

\bibitem{Sun2021}
Sun X, Wang L, Li H, et al. Identification of microenvironment related
  potential biomarkers of biochemical recurrence at 3 years after prostatectomy
  in prostate adenocarcinoma. {\it Aging (Albany NY)} 2021\string;
  13(12)\string: 16024-16042.
\newblock \href {\doibase 10.18632/aging.203121} {doi: 10.18632/aging.203121}

\bibitem{Chavda2022}
Chavda MP, Bihari S, Woodman RJ, Secombe P, Pilcher D. The impact of obesity on
  outcomes of patients admitted to intensive care after cardiac arrest. {\it J
  Crit Care} 2022\string; 69\string: 154025.
\newblock \href {\doibase 10.1016/j.jcrc.2022.154025} {doi:
  10.1016/j.jcrc.2022.154025}

\bibitem{Chen2022}
Chen C, Chen X, Chen J, et al. Association between Preoperative hs-crp/Albumin
  Ratio and Postoperative sirs in Elderly Patients: A Retrospective
  Observational Cohort Study. {\it J Nutr Health Aging} 2022\string;
  26(4)\string: 352-359.
\newblock \href {\doibase 10.1007/s12603-022-1761-4} {doi:
  10.1007/s12603-022-1761-4}

\bibitem{Ikeda2022a}
Ikeda T, Cooray U, Murakami M, Osaka K. Maintaining Moderate or Vigorous
  Exercise Reduces the Risk of Low Back Pain at 4 Years of Follow-Up: Evidence
  From the English Longitudinal Study of Ageing. {\it J Pain} 2022\string;
  23(3)\string: 390-397.
\newblock \href {\doibase 10.1016/j.jpain.2021.08.008} {doi:
  10.1016/j.jpain.2021.08.008}

\bibitem{Ikeda2022b}
Ikeda T, Tsuboya T. Effects of changes in depressive symptoms on handgrip
  strength in later life: A four-year longitudinal study in England. {\it J
  Affect Disord} 2022\string; 299\string: 67-72.
\newblock \href {\doibase 10.1016/j.jad.2021.11.057} {doi:
  10.1016/j.jad.2021.11.057}

\bibitem{MorenoBetancur2022}
Moreno-Betancur M, Lynch JW, Pilkington RM, et al. Emulating a target trial of
  intensive nurse home visiting in the policy-relevant population using linked
  administrative data. {\it Int J Epidemiol} 2022.
\newblock \href {\doibase 10.1093/ije/dyac092} {doi: 10.1093/ije/dyac092}

\bibitem{Figueroa2020EarlyApproaches}
Figueroa SC, Kennedy CJ, Wesseling C, Wiemels JM, Morimoto L, Mora AM. {Early
  immune stimulation and childhood acute lymphoblastic leukemia in Costa Rica:
  A comparison of statistical approaches}. {\it Environmental research}
  2020\string; 182\string: 109023.
\newblock \href {\doibase 10.1016/J.ENVRES.2019.109023} {doi:
  10.1016/J.ENVRES.2019.109023}

\bibitem{Pang2016a}
Pang M, Schuster T, Filion KB, Eberg M, Platt RW. Targeted Maximum Likelihood
  Estimation for Pharmacoepidemiologic Research. {\it Epidemiology}
  2016\string; 27(4)\string: 570-7.
\newblock \href {\doibase 10.1097/ede.0000000000000487} {doi:
  10.1097/ede.0000000000000487}

\bibitem{Pearl2016}
Pearl M, Balzer L, Ahern J. Targeted Estimation of Marginal Absolute and
  Relative Associations in Case-Control Data: An Application in Social
  Epidemiology. {\it Epidemiology} 2016\string; 27(4)\string: 512-7.
\newblock \href {\doibase 10.1097/ede.0000000000000476} {doi:
  10.1097/ede.0000000000000476}

\bibitem{Kreif2017}
Kreif N, Tran L, Grieve R, De~Stavola B, Tasker RC, Petersen M. Estimating the
  Comparative Effectiveness of Feeding Interventions in the Pediatric Intensive
  Care Unit: A Demonstration of Longitudinal Targeted Maximum Likelihood
  Estimation. {\it Am J Epidemiol} 2017\string; 186(12)\string: 1370-1379.
\newblock \href {\doibase 10.1093/aje/kwx213} {doi: 10.1093/aje/kwx213}

\bibitem{Almasi2018}
Almasi-Hashiani A, Nedjat S, Mansournia MA. Causal Methods for Observational
  Research: A Primer. {\it Arch Iran Med} 2018\string; 21(4)\string: 164-169.

\bibitem{Skeem2018ComparingIllness}
Skeem JL, Montoya L, Manchak SM. {Comparing Costs of Traditional and Specialty
  Probation for People With Serious Mental Illness}. {\it Psychiatric services
  (Washington, D.C.)} 2018\string; 69(8)\string: 896--902.
\newblock \href {\doibase 10.1176/APPI.PS.201700498} {doi:
  10.1176/APPI.PS.201700498}

\bibitem{Saddiki2018AScience}
Saddiki H, Balzer LB. {A Primer on Causality in Data Science}.  2018.
\newblock \href {\doibase 10.48550/arxiv.1809.02408} {doi:
  10.48550/arxiv.1809.02408}

\bibitem{Coyle2020TargetingResearch}
Coyle JR, Hejazi NS, Malenica I, et al. {Targeting Learning: Robust Statistics
  for Reproducible Research}.  2020.
\newblock \href {\doibase 10.48550/arXiv.2006.07333} {doi:
  10.48550/arXiv.2006.07333}

\bibitem{Gruber2012Tmle:Estimation}
Gruber S, \text{van der Laan} MJ. {tmle: An R Package for Targeted Maximum
  Likelihood Estimation}. {\it Journal of Statistical Software} 2012\string;
  51(13)\string: 1--35.
\newblock \href {\doibase https://doi.org/10.18637/jss.v051.i13} {doi:
  https://doi.org/10.18637/jss.v051.i13}

\bibitem{Lendle2017Ltmle:Data}
Lendle SD, Schwab J, Petersen ML, \text{van der Laan} MJ. {ltmle: An R Package
  Implementing Targeted Minimum Loss-Based Estimation for Longitudinal Data}.
  {\it Journal of Statistical Software} 2017\string; 81\string: 1--21.
\newblock \href {\doibase 10.18637/JSS.V081.I01} {doi: 10.18637/JSS.V081.I01}

\bibitem{Hernan2010TheRatios}
Hern{\'{a}}n MA. {The Hazards of Hazard Ratios}. {\it Epidemiology (Cambridge,
  Mass.)} 2010\string; 21(1)\string: 13.
\newblock \href {\doibase 10.1097/EDE.0B013E3181C1EA43} {doi:
  10.1097/EDE.0B013E3181C1EA43}

\bibitem{Phillips2022PracticalLearner}
Phillips RV, Laan v.~dMJ, Lee H, Gruber S. {Practical considerations for
  specifying a super learner}. {\it arXiv} 2022.
\newblock \href {\doibase doi.org/10.48550/arXiv.2204.06139} {doi:
  doi.org/10.48550/arXiv.2204.06139}

\bibitem{Moore2009}
Moore KL, \text{van der Laan} MJ. Covariate adjustment in randomized trials
  with binary outcomes: targeted maximum likelihood estimation. {\it Stat Med}
  2009\string; 28(1)\string: 39-64.
\newblock \href {\doibase 10.1002/sim.3445} {doi: 10.1002/sim.3445}

\bibitem{vanderLaan2010a}
\text{van der Laan} MJ. Targeted maximum likelihood based causal inference:
  Part I. {\it Int J Biostat} 2010\string; 6(2)\string: Article 2.
\newblock \href {\doibase 10.2202/1557-4679.1211} {doi: 10.2202/1557-4679.1211}

\bibitem{vanderLaan2010b}
\text{van der Laan} MJ. Targeted maximum likelihood based causal inference:
  Part II. {\it Int J Biostat} 2010\string; 6(2)\string: Article 3.
\newblock \href {\doibase 10.2202/1557-4679.1241} {doi: 10.2202/1557-4679.1241}

\bibitem{vanderLaan2012a}
\text{van der Laan} MJ, Balzer LB, Petersen ML. Adaptive matching in randomized
  trials and observational studies. {\it J Stat Res} 2012\string; 46(2)\string:
  113-156.

\bibitem{Schomaker2019}
Schomaker M, Luque-Fernandez MA, Leroy V, Davies MA. Using longitudinal
  targeted maximum likelihood estimation in complex settings with dynamic
  interventions. {\it Stat Med} 2019\string; 38(24)\string: 4888-4911.
\newblock \href {\doibase 10.1002/sim.8340} {doi: 10.1002/sim.8340}

\bibitem{Zheng2018}
Zheng W, Luo Z, \text{van der Laan} MJ. Marginal Structural Models with
  Counterfactual Effect Modifiers. {\it Int J Biostat} 2018\string; 14(1).
\newblock \href {\doibase 10.1515/ijb-2018-0039} {doi: 10.1515/ijb-2018-0039}

\bibitem{vanderLaan2012b}
\text{van der Laan} MJ, Gruber S. Targeted minimum loss based estimation of
  causal effects of multiple time point interventions. {\it Int J Biostat}
  2012\string; 8(1).
\newblock \href {\doibase 10.1515/1557-4679.1370} {doi: 10.1515/1557-4679.1370}

\bibitem{Schnitzer2016a}
Schnitzer ME, Lok JJ, Bosch RJ. Double robust and efficient estimation of a
  prognostic model for events in the presence of dependent censoring. {\it
  Biostatistics} 2016\string; 17(1)\string: 165-77.
\newblock \href {\doibase 10.1093/biostatistics/kxv028} {doi:
  10.1093/biostatistics/kxv028}

\bibitem{vanderLaan2014}
\text{van der Laan} MJ. Causal Inference for a Population of Causally Connected
  Units. {\it J Causal Inference} 2014\string; 2(1)\string: 13-74.
\newblock \href {\doibase 10.1515/jci-2013-0002} {doi: 10.1515/jci-2013-0002}

\bibitem{Sofrygin2017}
Sofrygin O, \text{van der Laan} MJ. Semi-Parametric Estimation and Inference
  for the Mean Outcome of the Single Time-Point Intervention in a Causally
  Connected Population. {\it J Causal Inference} 2017\string; 5(1).
\newblock \href {\doibase 10.1515/jci-2016-0003} {doi: 10.1515/jci-2016-0003}

\bibitem{Balzer2019}
Balzer LB, Zheng W, \text{van der Laan} MJ, Petersen ML. A new approach to
  hierarchical data analysis: Targeted maximum likelihood estimation for the
  causal effect of a cluster-level exposure. {\it Stat Methods Med Res}
  2019\string; 28(6)\string: 1761-1780.
\newblock \href {\doibase 10.1177/0962280218774936} {doi:
  10.1177/0962280218774936}

\bibitem{Balzer2021a}
Balzer LB, \text{van der Laan} M, Ayieko J, et al. Two-Stage TMLE to reduce
  bias and improve efficiency in cluster randomized trials. {\it Biostatistics}
  2021.
\newblock \href {\doibase 10.1093/biostatistics/kxab043} {doi:
  10.1093/biostatistics/kxab043}

\bibitem{Sofrygin2019}
Sofrygin O, Zhu Z, Schmittdiel JA, et al. Targeted learning with daily EHR
  data. {\it Stat Med} 2019\string; 38(16)\string: 3073-3090.
\newblock \href {\doibase 10.1002/sim.8164} {doi: 10.1002/sim.8164}

\bibitem{Gruber2013}
Gruber S, \text{van der Laan} MJ. An application of targeted maximum likelihood
  estimation to the meta-analysis of safety data. {\it Biometrics} 2013\string;
  69(1)\string: 254-62.
\newblock \href {\doibase 10.1111/j.1541-0420.2012.01829.x} {doi:
  10.1111/j.1541-0420.2012.01829.x}

\bibitem{Liu2022}
Liu Y, Schnitzer ME, Wang G, et al. Modeling treatment effect modification in
  multidrug-resistant tuberculosis in an individual patientdata meta-analysis.
  {\it Stat Methods Med Res} 2022\string; 31(4)\string: 689-705.
\newblock \href {\doibase 10.1177/09622802211046383} {doi:
  10.1177/09622802211046383}

\bibitem{Rose2009}
Rose S, Laan MJ. Why match? Investigating matched case-control study designs
  with causal effect estimation. {\it Int J Biostat} 2009\string; 5(1)\string:
  Article 1.
\newblock \href {\doibase 10.2202/1557-4679.1127} {doi: 10.2202/1557-4679.1127}

\bibitem{Rose2011}
Rose S, \text{van der Laan} MJ. A targeted maximum likelihood estimator for
  two-stage designs. {\it Int J Biostat} 2011\string; 7(1)\string: 17.
\newblock \href {\doibase 10.2202/1557-4679.1217} {doi: 10.2202/1557-4679.1217}

\bibitem{Balzer2015}
Balzer LB, Petersen ML, \text{van der Laan} MJ. Adaptive pair-matching in
  randomized trials with unbiased and efficient effect estimation. {\it Stat
  Med} 2015\string; 34(6)\string: 999-1011.
\newblock \href {\doibase 10.1002/sim.6380} {doi: 10.1002/sim.6380}

\bibitem{Balzer2016b}
Balzer LB, Petersen ML, \text{van der Laan} MJ. Targeted estimation and
  inference for the sample average treatment effect in trials with and without
  pair-matching. {\it Stat Med} 2016\string; 35(21)\string: 3717-32.
\newblock \href {\doibase 10.1002/sim.6965} {doi: 10.1002/sim.6965}

\bibitem{vanderLaan2010c}
\text{van der Laan} MJ, Gruber S. Collaborative double robust targeted maximum
  likelihood estimation. {\it Int J Biostat} 2010\string; 6(1)\string: Article
  17.
\newblock \href {\doibase 10.2202/1557-4679.1181} {doi: 10.2202/1557-4679.1181}

\bibitem{Pirracchio2018}
Pirracchio R, Yue JK, Manley GT, \text{van der Laan} MJ, Hubbard AE.
  Collaborative targeted maximum likelihood estimation for variable importance
  measure: Illustration for functional outcome prediction in mild traumatic
  brain injuries. {\it Stat Methods Med Res} 2018\string; 27(1)\string:
  286-297.
\newblock \href {\doibase 10.1177/0962280215627335} {doi:
  10.1177/0962280215627335}

\bibitem{Schnitzer2020}
Schnitzer ME, Sango J, Ferreira~Guerra S, \text{van der Laan} MJ. Data-adaptive
  longitudinal model selection in causal inference with collaborative targeted
  minimum loss-based estimation. {\it Biometrics} 2020\string; 76(1)\string:
  145-157.
\newblock \href {\doibase 10.1111/biom.13135} {doi: 10.1111/biom.13135}

\bibitem{Ju2019a}
Ju C, Gruber S, Lendle SD, et al. Scalable collaborative targeted learning for
  high-dimensional data. {\it Stat Methods Med Res} 2019\string; 28(2)\string:
  532-554.
\newblock \href {\doibase 10.1177/0962280217729845} {doi:
  10.1177/0962280217729845}

\bibitem{Ju2019b}
Ju C, Schwab J, \text{van der Laan} MJ. On adaptive propensity score truncation
  in causal inference. {\it Stat Methods Med Res} 2019\string; 28(6)\string:
  1741-1760.
\newblock \href {\doibase 10.1177/0962280218774817} {doi:
  10.1177/0962280218774817}

\bibitem{Ju2019c}
Ju C, Wyss R, Franklin JM, Schneeweiss S, Häggström J, \text{van der Laan}
  MJ. Collaborative-controlled LASSO for constructing propensity score-based
  estimators in high-dimensional data. {\it Stat Methods Med Res} 2019\string;
  28(4)\string: 1044-1063.
\newblock \href {\doibase 10.1177/0962280217744588} {doi:
  10.1177/0962280217744588}

\bibitem{Peterson2014}
Petersen M, Schwab J, Gruber S, Blaser N, Schomaker M, \text{van der Laan} MJ.
  Targeted Maximum Likelihood Estimation for Dynamic and Static Longitudinal
  Marginal Structural Working Models. {\it J Causal Inference} 2014\string;
  2(2)\string: 147-185.
\newblock \href {\doibase 10.1515/jci-2013-0007} {doi: 10.1515/jci-2013-0007}

\bibitem{Ferreira2020}
Ferreira~Guerra S, Schnitzer ME, Forget A, Blais L. Impact of discretization of
  the timeline for longitudinal causal inference methods. {\it Stat Med}
  2020\string; 39(27)\string: 4069-4085.
\newblock \href {\doibase 10.1002/sim.8710} {doi: 10.1002/sim.8710}

\bibitem{Zheng2016}
Zheng W, Petersen M, \text{van der Laan} MJ. Doubly Robust and Efficient
  Estimation of Marginal Structural Models for the Hazard Function. {\it Int J
  Biostat} 2016\string; 12(1)\string: 233-52.
\newblock \href {\doibase 10.1515/ijb-2015-0036} {doi: 10.1515/ijb-2015-0036}

\bibitem{VanDerLaan2016One-StepSubmodels}
\text{van der Laan} M, Gruber S. {One-Step Targeted Minimum Loss-based
  Estimation Based on Universal Least Favorable One-Dimensional Submodels}.
  {\it International Journal of Biostatistics} 2016\string; 12(1)\string:
  351--378.
\newblock \href {\doibase 10.1515/IJB-2015-0054} {doi: 10.1515/IJB-2015-0054}

\bibitem{Cai2020b}
Cai W, \text{van der Laan} MJ. One-step targeted maximum likelihood estimation
  for time-to-event outcomes. {\it Biometrics} 2020\string; 76(3)\string:
  722-733.
\newblock \href {\doibase 10.1111/biom.13172} {doi: 10.1111/biom.13172}

\bibitem{Zhu2020}
Zhu J, Gallego B. Targeted estimation of heterogeneous treatment effect in
  observational survival analysis. {\it J Biomed Inform} 2020\string;
  107\string: 103474.
\newblock \href {\doibase 10.1016/j.jbi.2020.103474} {doi:
  10.1016/j.jbi.2020.103474}

\bibitem{Rudolph2017}
Rudolph KE, \text{van der Laan} MJ. Robust estimation of encouragement-design
  intervention effects transported across sites. {\it J R Stat Soc Series B
  Stat Methodol} 2017\string; 79(5)\string: 1509-1525.
\newblock \href {\doibase 10.1111/rssb.12213} {doi: 10.1111/rssb.12213}

\bibitem{vanderLaan2015}
\text{van der Laan} MJ, Luedtke AR. Targeted Learning of the Mean Outcome under
  an Optimal Dynamic Treatment Rule. {\it J Causal Inference} 2015\string;
  3(1)\string: 61-95.
\newblock \href {\doibase 10.1515/jci-2013-0022} {doi: 10.1515/jci-2013-0022}

\bibitem{vanderLaan2022}
\text{van der Laan} L, Zhang W, Gilbert PB. Nonparametric estimation of the
  causal effect of a stochastic threshold-based intervention. {\it Biometrics}
  2022.
\newblock \href {\doibase 10.1111/biom.13690} {doi: 10.1111/biom.13690}

\bibitem{Zheng2012}
Zheng W, \text{van der Laan} MJ. Targeted maximum likelihood estimation of
  natural direct effects. {\it Int J Biostat} 2012\string; 8(1).
\newblock \href {\doibase 10.2202/1557-4679.1361} {doi: 10.2202/1557-4679.1361}

\bibitem{Lendle2013b}
Lendle SD, Subbaraman MS, \text{van der Laan} MJ. Identification and efficient
  estimation of the natural direct effect among the untreated. {\it Biometrics}
  2013\string; 69(2)\string: 310-7.
\newblock \href {\doibase 10.1111/biom.12022} {doi: 10.1111/biom.12022}

\bibitem{Rudolph2018}
Rudolph KE, Sofrygin O, Zheng W, \text{van der Laan} MJ. Robust and Flexible
  Estimation of Stochastic Mediation Effects: A Proposed Method and Example in
  a Randomized Trial Setting. {\it Epidemiol Methods} 2018\string; 7(1).
\newblock \href {\doibase 10.1515/em-2017-0007} {doi: 10.1515/em-2017-0007}

\bibitem{Box1976ScienceStatistics}
Box GE. {Science and statistics}. {\it Journal of the American Statistical
  Association} 1976\string; 71(356)\string: 791--799.
\newblock \href {\doibase 10.1080/01621459.1976.10480949} {doi:
  10.1080/01621459.1976.10480949}

\bibitem{Chernozhukov2018Double/debiasedParameters}
Chernozhukov V, Chetverikov D, Demirer M, et al. {Double/debiased machine
  learning for treatment and structural parameters}. {\it The Econometrics
  Journal} 2018\string; 21(1)\string: C1-C68.
\newblock \href {\doibase 10.1111/ECTJ.12097} {doi: 10.1111/ECTJ.12097}

\bibitem{Zivich2021}
Zivich PN, Breskin A. Machine Learning for Causal Inference: On the Use of
  Cross-fit Estimators. {\it Epidemiology} 2021\string; 32(3)\string: 393-401.
\newblock \href {\doibase 10.1097/ede.0000000000001332} {doi:
  10.1097/ede.0000000000001332}

\bibitem{Gruber2010b}
Gruber S, \text{van der Laan} MJ. A targeted maximum likelihood estimator of a
  causal effect on a bounded continuous outcome. {\it Int J Biostat}
  2010\string; 6(1)\string: Article 26.
\newblock \href {\doibase 10.2202/1557-4679.1260} {doi: 10.2202/1557-4679.1260}

\bibitem{Balzer2016a}
Balzer L, Ahern J, Galea S, \text{van der Laan} M. Estimating Effects with Rare
  Outcomes and High Dimensional Covariates: Knowledge is Power. {\it Epidemiol
  Methods} 2016\string; 5(1)\string: 1-18.
\newblock \href {\doibase 10.1515/em-2014-0020} {doi: 10.1515/em-2014-0020}

\bibitem{Benkeser2018}
Benkeser D, Carone M, Gilbert PB. Improved estimation of the cumulative
  incidence of rare outcomes. {\it Stat Med} 2018\string; 37(2)\string:
  280-293.
\newblock \href {\doibase 10.1002/sim.7337} {doi: 10.1002/sim.7337}

\bibitem{Diaz2016}
Díaz I, Colantuoni E, Rosenblum M. Enhanced precision in the analysis of
  randomized trials with ordinal outcomes. {\it Biometrics} 2016\string;
  72(2)\string: 422-31.
\newblock \href {\doibase 10.1111/biom.12450} {doi: 10.1111/biom.12450}

\bibitem{Diaz2021NonparametricPolicies}
D{\'{i}}az I, Williams N, Hoffman KL, Schenck EJ. {Nonparametric Causal Effects
  Based on Longitudinal Modified Treatment Policies}. {\it Journal of the
  American Statistical Association} 2021.
\newblock \href {\doibase 10.1080/01621459.2021.1955691} {doi:
  10.1080/01621459.2021.1955691}

\bibitem{Diaz2017}
Díaz I, \text{van der Laan} MJ. Doubly robust inference for targeted minimum
  loss-based estimation in randomized trials with missing outcome data. {\it
  Stat Med} 2017\string; 36(24)\string: 3807-3819.
\newblock \href {\doibase 10.1002/sim.7389} {doi: 10.1002/sim.7389}

\bibitem{grubertmle}
Gruber S, {van der Laan} MJ. {tmle}: An {R} Package for Targeted Maximum
  Likelihood Estimation. {\it Journal of Statistical Software} 2012\string;
  51(13)\string: 1--35.

\bibitem{benkeser2017improved}
Benkeser DC, Carone M, Gilbert PB. Improved estimation of the cumulative
  incidence of rare outcomes. {\it Statistics in Medicine} 2017.
\newblock \href {\doibase 10.1002/sim.7337} {doi: 10.1002/sim.7337}

\bibitem{benkeser2017survtmle}
Benkeser DC, Hejazi NS. {survtmle}: Targeted Minimum Loss-Based Estimation for
  Survival Analysis in {R}.  2017.
\newblock \href {\doibase 10.5281/zenodo.835868} {doi: 10.5281/zenodo.835868}

\bibitem{drtmlepackage}
Benkeser D. drtmle: Doubly-Robust Nonparametric Estimation and Inference.
  2017.
\newblock \href {\doibase 10.5281/zenodo.844836} {doi: 10.5281/zenodo.844836}

\bibitem{sofrygin2017stremr}
Sofrygin O, {\text{van der Laan}} MJ, Neugebaueer R. stremr: Streamlined
  Estimation for Static, Dynamic and Stochastic Treatment Regimes in
  Longitudinal Data. {\it R package} 2017.
\newblock \href {\doibase https://github.com/osofr/stremr} {doi:
  https://github.com/osofr/stremr}

\bibitem{coyle2018origami}
Coyle JR, Hejazi NS. origami: A Generalized Framework for Cross-Validation in
  R. {\it The Journal of Open Source Software} 2018\string; 3(21).
\newblock \href {\doibase 10.21105/joss.00512} {doi: 10.21105/joss.00512}

\bibitem{coyle2022hal9001-rpkg}
Coyle JR, Hejazi NS, Phillips RV, {van der Laan} L, {van der Laan} MJ.
  {hal9001}: The scalable highly adaptive lasso.
  https://doi.org/10.5281/zenodo.3558313;  2022.
\newblock {R} package version 0.4.2

\bibitem{hejazi2020hal9001-joss}
Hejazi NS, Coyle JR, {van der Laan} MJ. {hal9001}: Scalable highly adaptive
  lasso regression in {R}. {\it Journal of Open Source Software} 2020.
\newblock \href {\doibase 10.21105/joss.02526} {doi: 10.21105/joss.02526}

\bibitem{coyle2021sl3-rpkg}
Coyle JR, Hejazi NS, Malenica I, Phillips RV, Sofrygin O. {sl3}: Modern
  Pipelines for Machine Learning and {Super Learning}.
  \url{https://github.com/tlverse/sl3};  2021.
\newblock {R} package version 1.4.2

\bibitem{coyle2021tmle3-rpkg}
Coyle JR. {tmle3}: The Extensible {TMLE} Framework.  2021.
\newblock \href {\doibase 10.5281/zenodo.4603358} {doi: 10.5281/zenodo.4603358}

\bibitem{malenica2022tmle3mopttx}
Malenica I, Coyle J, {van der Laan} MJ. {tmle3mopttx}: Targeted Learning and
  Variable Importance with Optimal Individualized Categorical Treatment.
  https://github.com/tlverse/tmle3mopttx;  2022.
\newblock R package version 1.0.0

\bibitem{hejazi2021tmle3shift-rpkg}
Hejazi NS, Coyle JR, {van der Laan} MJ. {tmle3shift}: {Targeted Learning} of
  the Causal Effects of Stochastic Interventions.
  \url{https://github.com/tlverse/tmle3shift};  2021.
\newblock {R} package version 0.2.0

\bibitem{ctmle2017}
Ju C, Gruber S, Laan v.~dM. ctmle: Collaborative Targeted Maximum Likelihood
  Estimation.  2017.
\newblock \href {\doibase https://CRAN.R-project.org/package=ctmle} {doi:
  https://CRAN.R-project.org/package=ctmle}

\bibitem{Hejazi2022Haldensify:R}
Hejazi NS, \text{van der Laan} MJ, Benkeser D. {haldensify: Highly adaptive
  lasso conditional density estimation in R}. {\it Journal of Open Source
  Software} 2022\string; 7(77)\string: 4522.
\newblock \href {\doibase 10.21105/JOSS.04522} {doi: 10.21105/JOSS.04522}

\bibitem{Ertefaie2020NonparametricLasso}
Ertefaie A, Hejazi NS, \text{van der Laan} MJ. {Nonparametric inverse
  probability weighted estimators based on the highly adaptive lasso}. {\it
  Biometrics} 2022.
\newblock \href {\doibase 10.1111/biom.13719} {doi: 10.1111/biom.13719}

\bibitem{hejazi2020efficient}
Hejazi NS, {\text{van der Laan}} MJ, Janes HE, Gilbert PB, Benkeser DC.
  Efficient nonparametric inference on the effects of stochastic interventions
  under two-phase sampling, with applications to vaccine efficacy trials. {\it
  Biometrics} 2020.
\newblock \href {\doibase 10.1111/biom.13375} {doi: 10.1111/biom.13375}

\bibitem{hejazi2020txshift-joss}
Hejazi NS, Benkeser DC. {txshift}: Efficient estimation of the causal effects
  of stochastic interventions in {R}. {\it Journal of Open Source Software}
  2020.
\newblock \href {\doibase 10.21105/joss.02447} {doi: 10.21105/joss.02447}

\bibitem{hejazi2022txshift-rpkg}
Hejazi NS, Benkeser DC. {txshift}: Efficient Estimation of the Causal Effects
  of Stochastic Interventions. {\it Journal of Open Source Software} 2022.
\newblock \href {\doibase 10.5281/zenodo.4070042} {doi: 10.5281/zenodo.4070042}

\bibitem{zepid}
Zivich P, Davidson-Pilon C, Diong J, Reger D, Badger TG. zEpid.
  \url{https://github.com/pzivich/zepid};  2022.
\newblock {Python} package version 0.9.1

\bibitem{Luque-Fernandez2021ELTMLE:Estimation}
Luque-Fernandez MA. {ELTMLE: Stata module to provide Ensemble Learning Targeted
  Maximum Likelihood Estimation}. {\it Statistical Software Components} 2021.
\newblock \href {\doibase https://ideas.repec.org/c/boc/bocode/s458337.html}
  {doi: https://ideas.repec.org/c/boc/bocode/s458337.html}

\bibitem{GruberArxiv}
Gruber S, Phillips RA, Lee H, Ho M, Concato J, \text{van der Laan} MJ. Targeted
  learning: Towards a future informed by real-world evidence. {\it Arxiv} 2022.
\newblock \href {\doibase arXiv:2205.08643} {doi: arXiv:2205.08643}

\bibitem{SentinelInnov}
Sentinel Innovation~Center . Sentinel Innovation Center Master Plan.  2021.
\newblock \href {\doibase
  https://www.sentinelinitiative.org/news-events/publications-presentations/innovation-center-ic-master-plan}
  {doi:
  https://www.sentinelinitiative.org/news-events/publications-presentations/innovation-center-ic-master-plan}

\bibitem{gruber2022developing}
Gruber S, Lee H, Phillips R, Ho M, Laan v.~dM. Developing a Targeted
  Learning-Based Statistical Analysis Plan. {\it Statistics in
  Biopharmaceutical Research} 2022\string: 1--8.
\newblock \href {\doibase https://doi.org/10.1080/19466315.2022.2116104} {doi:
  https://doi.org/10.1080/19466315.2022.2116104}

\end{thebibliography}


\begin{thebibliography}{10}

\bibitem{Hirt1974}
Hirt CW, Amsden AA, Cook JL. An arbitrary {L}agrangian-{E}ulerian computing
  method for all flow speeds.  {\it J {C}omput {P}hys. }1974;14(3):227--253.

\bibitem{Liska2010}
Liska R, Shashkov M, Vachal P, Wendroff B. Optimization-based synchronized
  flux-corrected conservative interpolation (remapping) of mass and momentum
  for arbitrary {L}agrangian-{E}ulerian methods.  {\it J {C}omput {P}hys.
  }2010;229(5):1467--1497.

\bibitem{Taylor1937}
Taylor GI, Green AE. Mechanism of the production of small eddies from large
  ones.  {\it P {R}oy {S}oc {L}ond {A} {M}at. }1937;158(895):499--521.
\newblock \url{https://doi.org/10.1098/rspa.1937.0036},
  \url{http://rspa.royalsocietypublishing.org/content/158/895/499}.

\bibitem{Knupp1999}
Knupp PM. Winslow smoothing on two-dimensional unstructured meshes.  {\it Eng
  {C}omput. }1999;15:263--268.

\bibitem{Kamm2000}
Kamm J. {\it Evaluation of the {S}edov-von {N}eumann-{T}aylor blast wave
  solution. } Technical {R}eport LA-UR-00-6055: Los {A}lamos {N}ational
  {L}aboratory; 2000.

\bibitem{Kucharik2003}
Kucharik M, Shashkov M, Wendroff B. An efficient linearity-and-bound-preserving
  remapping method.  {\it J {C}omput {P}hys. }2003;188(2):462--471.

\bibitem{Blanchard2015}
Blanchard G, Loubere R. {\it High-Order {C}onservative {R}emapping with a
  posteriori {MOOD} stabilization on polygonal meshes. }
  \url{https://hal.archives-ouvertes.fr/hal-01207156}, the {HAL} {O}pen
  {A}rchive, hal-01207156. Accessed January 13, 2016; 2015.

\bibitem{Burton2013}
Burton DE, Kenamond MA, Morgan NR, Carney TC, Shashkov MJ. An intersection
  based {ALE} scheme {(xALE)} for cell centered hydrodynamics {(CCH)}.  In:
  Talk at {M}ultimat 2013, {I}nternational {C}onference on {N}umerical
  {M}ethods for {M}ulti-{M}aterial {F}luid {F}lows; September 2--6, 2013; San
  {F}rancisco.
\newblock LA-UR-13-26756.2.

\bibitem{Berndt2011}
Berndt M, Breil J, Galera S, Kucharik M, Maire PH, Shashkov M. Two-step hybrid
  conservative remapping for multimaterial arbitrary {L}agrangian-{E}ulerian
  methods.  {\it J {C}omput {P}hys. }2011;230(17):6664--6687.

\bibitem{Kucharik2012}
Kucharik M, Shashkov M. One-step hybrid remapping algorithm for multi-material
  arbitrary {L}agrangian-{E}ulerian methods.  {\it J {C}omput {P}hys.
  }2012;231(7):2851--2864.

\bibitem{Breil2015}
Breil J, Alcin H, Maire PH. A swept intersection-based remapping method for
  axisymmetric {ReALE} computation.  {\it Int {J} {N}umer {M}eth {F}l.
  }2015;77(11):694--706.
\newblock Fld.3996.

\bibitem{Barth1997}
Barth TJ. Numerical methods for gasdynamic systems on unstructured meshes.  In:
   Kroner D, Rohde C, Ohlberger M, eds. {\it An {I}ntroduction to {R}ecent
  {D}evelopments in {T}heory and {N}umerics for {C}onservation {L}aws,
  {P}roceedings of the {I}nternational {S}chool on {T}heory and {N}umerics for
  {C}onservation {L}aws}, Lecture {N}otes in {C}omputational {S}cience and
  {E}ngineering. Berlin: Springer 1997.
\newblock ISBN 3-540-65081-4.

\bibitem{Lauritzen2011}
Lauritzen P, Erath C, Mittal R. On simplifying `incremental remap'-based
  transport schemes.  {\it J {C}omput {P}hys. }2011;230(22):7957--7963.

\bibitem{Klima2017}
Klima M, Kucharik M, Shashkov M. Local error analysis and comparison of the
  swept- and intersection-based remapping methods.  {\it Commun {C}omput
  {P}hys. }2017;21(2):526--558.

\bibitem{Dukowicz2000}
Dukowicz JK, Baumgardner JR. Incremental remapping as a transport/advection
  algorithm.  {\it J {C}omput {P}hys. }2000;160(1):318--335.

\bibitem{Kucharik2011}
Kucharik M, Shashkov M. Flux-based approach for conservative remap of
  multi-material quantities in {2D} arbitrary {L}agrangian-{E}ulerian
  simulations.  In:  Fo\v{r}t J, F{\"{u}}rst J, Halama J, Herbin R, Hubert F,
  eds. {\it Finite {V}olumes for {C}omplex {A}pplications {VI} {P}roblems \&
  {P}erspectives},  Springer {P}roceedings in {M}athematics, vol. 1: Springer
  2011 (pp. 623--631).

\bibitem{Kucharik2014}
Kucharik M, Shashkov M. Conservative multi-material remap for staggered
  multi-material arbitrary {L}agrangian-{E}ulerian methods.  {\it J {C}omput
  {P}hys. }2014;258:268--304.

\bibitem{Loubere2005}
Loubere R, Shashkov M. A subcell remapping method on staggered polygonal grids
  for arbitrary-{L}agrangian-{E}ulerian methods.  {\it J {C}omput {P}hys.
  }2005;209(1):105--138.

\bibitem{Caramana1998}
Caramana EJ, Shashkov MJ. Elimination of artificial grid distortion and
  hourglass-type motions by means of {L}agrangian subzonal masses and
  pressures.  {\it J {C}omput {P}hys. }1998;142(2):521--561.

\bibitem{Hoch2009}
Hoch P. {\it An arbitrary {L}agrangian-{E}ulerian strategy to solve
  compressible fluid flows. } Technical {R}eport: CEA; 2009.
\newblock HAL: hal-00366858.
  https://hal.archives-ouvertes.fr/docs/00/36/68/58/PDF/ale2d.pdf. Accessed
  January 13, 2016.

\bibitem{Shashkov1996}
Shashkov M. {\it Conservative {F}inite-{D}ifference {M}ethods on {G}eneral
  {G}rids}.
\newblock Boca Raton, Florida: CRC {P}ress; 1996.
\newblock ISBN 0-8493-7375-1.

\bibitem{Benson1992}
Benson DJ. Computational methods in {L}agrangian and {E}ulerian hydrocodes.
  {\it Comput {M}ethod {A}ppl {M}. }1992;99(2--3):235--394.

\bibitem{Margolin2003}
Margolin LG, Shashkov M. Second-order sign-preserving conservative
  interpolation (remapping) on general grids.  {\it J {C}omput {P}hys.
  }2003;184(1):266--298.

\bibitem{Kenamond2013}
Kenamond MA, Burton DE. Exact intersection remapping of multi-material
  domain-decomposed polygonal meshes.  In: Talk at {M}ultimat 2013,
  {I}nternational {C}onference on {N}umerical {M}ethods for {M}ulti-{M}aterial
  {F}luid {F}lows; September 2--6, 2013; San {F}rancisco.
\newblock LA-UR-13-26794.

\bibitem{Dukowicz1984}
Dukowicz J. Conservative rezoning (remapping) for general quadrilateral meshes.
   {\it J {C}omput {P}hys. }1984;54(3):411--424.

\bibitem{Margolin2002}
Margolin LG, Shashkov M. {\it Second-order sign-preserving remapping on general
  grids. } Technical Report LA-UR-02-525: Los {A}lamos {N}ational {L}aboratory;
  2002.

\bibitem{Mavriplis2003}
Mavriplis DJ. Revisiting the least-squares procedure for gradient
  reconstruction on unstructured meshes.  In: AIAA 2003-3986. 16th {AIAA}
  {C}omputational {F}luid {D}ynamics {C}onference; June 23--26, 2003; Orlando,
  {F}lorida.

\bibitem{Scovazzi2008}
Scovazzi G, Love E, Shashkov M. Multi-scale {L}agrangian shock hydrodynamics on
  {Q1/P0} finite elements: {T}heoretical framework and two-dimensional
  computations.  {\it Comput {M}ethod {A}ppl {M}. }2008;197(9--12):1056--1079.

\end{thebibliography}

\newpage

\appendix
        
    \begin{table}[h!]
    \centering
    \caption{TMLE methodological developments and their associated extensions}
    \label{tab:Table3}
    \begin{tabular}{llll}
    \hline
    Year & Development & First author & Related developments \\ \hline
    2006 & Seminal paper (TMLE) & van der Laan MJ &  \\ \hline
    2009 & Small sample size & Moore KL & Gruber S, 2010 \\ \hdashline
     & Case-control studies & Rose S & Rose S, 2011 \\
     &  &  & Balzer L, 2015 \\
     &  &  & Balzer L, 2016 \\ \hline
    2010 & Collaborative TMLE & van der Laan MJ & Pirracchio R, 2018 \\
     & (c-TMLE) &  & Ju C, 2019 \\ \hdashline
     & Time to event data &  & Ju C, 2019 \\
     &  &  & Ju C, 2019 \\
     &  &  & Schnitzer ME, 2020 \\  \hdashline
     & Longitudinal   data (LTMLE) & van der Laan MJ & van der Laan, 2012 \\
     &  &  & Schomaker M, 2019 \\  \hline
    2012 & Sequential randomised trials & Chaffee PH &  \\  \hdashline
     & Natural direct effect & Zheng W & Lendle SD, 2013 \\
     &  &  & Rudolph KE, 2018 \\  \hdashline
     & Non-independence & van der Laan MJ & van der Laan MJ, 2014 \\
     &  &  & Schnitzer M, 2016 \\
     &  &  & Sofrygin O, 2017 \\
     &  &  & Balzer L, 2019 \\
     &  &  & Balzer L, 2021 \\ \hline
    2013 & Meta-analysis / Safety outcomes & Gruber S & Liu Y, 2022 \\  \hline
    2014 & Pooled TMLE & Petersen M & Zheng W, 2016 \\
     &  &  & Ferreira Guerra S, 2020 \\  \hdashline
     & Interval-censored TMLE & Sapp S &  \\  \hdashline
     & Genetics & Wang H & Benkeser D, 2019 \\
     &  &  & Yang G, 2022 \\  \hline
    2015 & PS & Lendle SD &  \\
     & Cross-validated TMLE & van der Laan MJ &  \\  \hdashline
     & cv-TMLE &  &  \\  \hline
    2016 & One-step TMLE & van der Laan MJ & Cai W, 2020 \\
     &  &  & Zhu J, 2020 \\
     & TMLE for rare outcomes & Balzer L & Benkeser D, 2018 \\  \hdashline
     & TMLE for ordinal outcomes & Díaz I &  \\  \hline
    2017 & TMLE with missing outcome data & Díaz I &  \\  \hdashline
     & Robust TMLE & Rudolph KE &  \\  \hdashline
     & Targeted sequential inference of an optimal treatment rule & Chambaz A &  \\  \hline
    2018 & Projected TMLE & Zheng W &  \\  \hline
    2019 & TMLE for cluster-level exposure & Balzer LB &  \\  \hdashline
     & Long-format TMLE & Sofrygin O &  \\  \hline
    2020 & \begin{tabular}[c]{@{}l@{}}Highly-Adaptive least absolute shrinkage and selection \\ operator (LASSO) Targeted Minimum Loss Estimator \\ (HAL-TMLE)\end{tabular} & Cai W &  \\  \hline
    2022 & Threshold response function & van der Laan MJ &  \\ \hline
    \end{tabular}
    \end{table}
    
    \newpage \,
    \newpage


    \begin{landscape}
       

\section*{Supplementary material}

**TABLE 2** Articles by discipline

| Authors | Year | Journal | Disease area | Research question | Challenges | Contribution to research |
|---|---|---|---|---|---|---|
| **Behavioural epidemiology** | | | | | | |
| Mackey DC et al. | 2011 | Am J Epidemiol | Healthy living<br>Older age | Physical activity and hip fracture | No RCTs | Little difference in hip fracture risk for men with moderate or high physical activity levels relative to those with low physical activity level. |
| Ahern J et al. | 2016 | Epidemiology | Mental health<br>Childhood hardship<br>Adolescent health<br>Ethnicity | Childhood adversities and mental disorders in adolescents | Multidimensional exposure | Childhood adversities play an important role in the burden of mental disorders in adolescents, particularly for behavior disorders and to some extent for distress and substance disorders.<br><br>However, despite the substantially higher burden of adversities experienced by black and Hispanic adolescents, they have a minor role in the patterns of disparities between racial/ethnic groups in mental disorders. Notably, fear disorders were the most common in all racial/ethnic groups, highest in black youth, and largely unrelated to childhood adversities |
| Platt JM et al. | 2018 | Am J Epidemiol | Childhood hardship<br>IQ | Association between 11 childhood adversities and intelligence | Multidimensional exposure | Interventions attempting to support and improve cognition in individuals who report childhood adversity can be a useful complement to interventions for emotional and behavioral disturbances |
| Rodríguez-Molina D et al. | 2019 | Int Arch Occup Environ Health | Occupational epi<br>Respiratory disease | Improve knowldedge about prevention against occupational asthma | Treatment effect heterogeneity bias<br>Low sample size<br>High prop missing values | We found that using an instructional video as educational intervention is an effective approach to improve knowledge about preventive measures against occupational asthma and allergies in Bavarian farm apprentices. |
| Torres JM et al. | 2019 | Epidemiology | Migration<br>Older age<br>Care<br>Deprivation | Is family-member migration associated with unmet caregiving needs among older adults who remain in low and middle-income settings? | Longitudinal relationships<br>Sample attrition<br>Time-varying covariates | It may be that the long-term – but not short-term – absence of adult children had adverse consequences for women's physical functioning as they aged into older adulthood.<br>We also found entirely null associations between having an adult child in the US and physical functioning for men. |
| Ehrlich SF et al. | 2020 | Am J Epidemiol | Pregnancy / prenatal exposure<br>Healthy living<br>Birth outcomes | Risk of small or large for gestational age according to exercise during 1st trimester of pregnancy | Discussion of assumptions<br>Model mis-specification | In underweight and normal-weight women only, meeting the lower exercise threshold recommended by the Physical Activity Guidelines for Americans also appears to increase the risk of SGA and decrease the risk of LGA |
| Bodnar LM et al. | 2020 | Am J Clin Nutr | Nutritional epi<br>Pregnancy | Associations between fruit and vegetable intake relative to total energy intake and adverse pregnancy outcomes | Dichotomisation of exposure<br>Complex interactive effects between exposure and outcome<br>Multidimensional exposure<br>Curse of dimensionality | TMLE produced effect estimates with less variation that suggested protective associations for diets high in fruits and vegetables relative to energy on risk of preterm birth, SGA birth, and pre-eclampsia. |

| | | | | | | |
|---|---|---|---|---|---|---|
| Kagawa RMC et al. | 2020 | Ann Epidemiol | Social epi<br>Deprivation<br>Mental health | Effect of fire arm involvement during violent victimization on the level of distress experienced and daily functioning within sociodemographic subgroups. | Missing data | Victimization with a firearm is more distressing than victimization with another weapon or no weapon and that this response is almost universal across age, sex, race, and socioeconomic position. |
| Puryear SB et al. | 2020 | AIDS | HIV<br>Alcohol | Assessing the effect of alcohol use across the entire cascade, from diagnosis to viral suppression. | Longitudinal effects | Via the multiple steps of the cascade, HIV-positive drinkers had significantly worse viral suppression outcomes than non-drinkers. |
| Torres JM et al. | 2020 | Am J Epidemiol | Migration<br>Older age<br>Physical activity | Association between adult child US migration status and change in cognitive performance scores | | |
| Clare PJ et al. | 2020 | Addiction | Alcohol<br>Adolescent health | Effect of parental supply of alcohol on alcohol-related outcomes in early adulthood | Time-varying confounding<br>Causal effects | Further evidence that parental supply of alcohol in adolescence has effects on a number of negative outcomes in early adulthood, including binge drinking and alcohol-related harm, leading not only to increased risk of binge drinking and harm but also increased frequency of binge drinking and number of harms experienced. Analysis of earlier initiation of supply showed that the magnitude of the effect of parental supply increased the earlier that supply was initiated |
| Kang L et al. | 2021 | J Safety Res | | Risk-taking behaviors under various urban-street conditions, as a function of helmet use | Self selection<br>Survey<br>Heterogeneous effects<br>Model mis-specification | Based on 131 survey participants, a significant positive risk compensation effect has been identified using the TMLE estimator and the size of effect is estimated to be about 15.6%. |
| Torres JM et al. | 2021 | Int J Geriatr Psychiatry | Older age<br>Caring responsibilities<br>Health outcomes | Evaluated the effect of spousal caregiving on multiple health outcomes in middle-aged and older adults in Mexico. | Reverse causality<br>Survey waves<br>Longitudinal data | Select evidence of adverse associations between spousal caregiving and past-week depressive symptoms: These adverse associations are generally described as the result of the emotional and physical burden of caregiving, which may have negative consequences for sleep, time for leisure and health promoting activities, and social isolation. |
| Lee JO et al. | 2021 | Public Health | Employment<br>Mental health<br>Covid-19 | Examined the association of employment insecurity with two mental health measures, depression and anxiety | Reverse causality<br>Missing data<br>Sample weights | Employment insecurity has threatened mental health in the United States during the pandemic, and mental health repercussions are not felt equally across the population. |

| Author | Year | Journal | Topic | Aim | Limitations | Findings |
|---|---|---|---|---|---|---|
| Shiba K et al. | 2021 | Epidemiology | Older age<br>Mental health | We present an analysis that estimates and compares prevalence of depressive symptoms under alternattive hypothetical interventions in social participation. | Longitudinal data<br>Survey waves<br>Reverse causation<br>Measurement bias/mis-classification | Past studies linking social participation and depressive symptoms in late life have not rigorously considered the time-varying nature of social participation.<br>First study that explicitly estimated and compared the effects of alternative hypothetical interventions in social participation at two time points on subsequent depressive symptoms. |
| Ikeda T et al. | 2022 | J Affect Disord | Older age<br>Mental health<br>Strength | Examine the magnitude of the association between depressive symptoms over 2 years and weak handgrip strength among English people in 4 years of follow-up. | Survey waves<br>Sample attrition - selection bias<br>Self reported exposure | The main finding of our study was that people who maintained non-depressive symptoms or improved depressive symptoms were less likely to have weak handgrip strength than those with persistent depressive symptoms. |
| Ikeda T et al. | 2022 | J Pain | Older age<br>Physical activity | We hypothesized that older individuals who maintained physical activity over time tend to have a lower risk of low back pain, whereas those who discontinued activity were not. | Self reported exposure<br>Small sample<br>Sample attrition | Overall, the present study confirmed that maintaining physical activity reduced the risk of low back pain at the follow-up survey. Conversely, discontinuing activity (engaged only at the baseline survey) was not beneficial. |
| **Biomarker epidemiology** | | | | | | |
| Bembom O et al. | 2009 | Stat Med | Mutations<br>Virology | Determine which of a set of candidate viral mutations affects clinical virologic response to the antiretroviral drug lopinavir + rank the importance of these mutations for drug-specific resistance | Identify subset of relevant biomarkers<br>Biomarker importance | The subset of mutations identified by this approach as significant contributors to lopinavir resistance was in better agreement with the current knowledge than the subsets identified by an unadjusted analyses or the G-computation approach. In addition, the specific ranking provided by targeted VIM estimation also agreed better with the current understanding than did the rankings generated with alternative methods. |
| Rosenblum M et al. | 2009 | PLoS One | HIV<br>Virology | Effect of adherence on viral load after different durations of viral suppression. | Selection bias<br>Unmeasured confounding | These data suggest that for adherence proportions greater than 50%, the probability of virologic failure decreases with longer duration of viral suppression. |
| Gianfrancesco MA et al. | 2016 | Genes Immun | Genetics<br>Ethnicity<br>Skin disease | Examine clinically important marginal effects of a Genetic Risk Score composed of 41 established genetic risk loci on systemic lupus erythematosus activity over a period of 9 years | Little awareness of how genetic profiles have an impact on disease activity<br>Longitudinal effects<br>Time-dependent confounding<br>Sample attrition | Results from individual SNP analyses provide important insight to the overall GRS findings; specifically, evidence for significant associations between certain SNPs and SLAQ score at two time points during the longitudinal study was demonstrated. |

| Author | Year | Journal | Topic | Objective | Methodological challenges | Main findings |
|---|---|---|---|---|---|---|
| Hsu LI et al. | 2016 | Cancer Epidemiol Biomarkers Prev | Ethnicity Cancer Children health | Identify a list of candidate genes within each significantly enriched pathway in childhood Leukemia, while accounting in models, for the complex correlation between SNPs | Large, correlated data Data reduction Gene selection Variable importance | The results demonstrate that newly developed bioinformatics tools and causal inference methods may illuminate new and biologically relevant pathways and genes to improve current understanding of pathogenesis in childhood leukemia. |
| Salihu HM et al. | 2016 | Matern Child Health J | Ethnicity Pregnancy outcomes | Describe the methylation patterns of 20 candidate genes associated with preterm birth and evaluate their role in preterm births in African-American women. | Residual confounding | This study examined 42 CpG sites within 20 candidate genes previously linked to preterm birth and identified three CpG sites on 2 distinct genes (TNF-a and PON1) that were differentially methylated between black and nonblack newborns. |
| Izano MA et al. | 2020 | PLoS One | Birth Mental health | Evaluate the relationship between newborn telomere length and a comprehensive suite of chronic maternal stressors | Complex joint effects | We found that a greater proportion of Latina mothers reported financial strain, food insecurity, and high job strain, while a greater proportion of Black mothers reported poor neighborhood quality, experiencing stressful/traumatic life events, or having an unplanned pregnancy than other racial/ethnic groups. |
| Wang L et al. | 2021 | Front Genet | Cancer Environmental epidemiology | Identify tumor microenvironment-related genes to estimate their effects on the 3-year mortality of Ovarian cancer. | Variable selection Complex joint effects Model mis-specification | ARID3C, CROCC2, FREM2, and PTF1A were identified as prognostic biomarkers for OSC patients. Two of them (FREM2 and PTF1A), alongside CROCC, were successfully validated in three GEO datasets. |
| Sun X et al. | 2021 | Aging (Albany NY) | Cancer | Identify the potential prognostic genes in the prostate adenocarcinoma microenvironment and estimate the causal effects simultaneously. | Causal effects Variable selection | Based on this strategy, we identified 14 genes involved in the prognosis of Prostate adenocarcinoma. The interaction between PRAD and TME might have serious effects on tumor evolution, further influencing tumor resistance, recurrence, and overall prognosis. |
| **Environmental epidemiology** | | | | | | |
| Padula AM et al. | 2012 | Am J Epidemiol | Prenatal exposure Birth outcomes | Estimate, at the population level, the predicted probability of term LBW had everyone been exposed to each quartile of traffic density. | Clear definitions of outcome and exposure + Previous inconsistent findings Measurement error Misclassification of exposure | The results did not show a clear exposure-response relation across the quartiles of traffic density; however, there was a significant difference in the predicted probability of LBW between the highest and lowest quartiles of exposure, showing that higher traffic density is associated with increased probability of LBW. |

| Author | Year | Journal | Topic | Objective | Methods/Issues | Findings |
|---|---|---|---|---|---|---|
| Herrera R et al. | 2017 | Int J Environ Res Public Health | Respiratory conditions Children health | Quantify the causal attributable risk of living close to the mines on asthma or allergic rhinoconjunctivitis risk burden in children | Assumptions discussed/checked Proxy measures of exposure Questionnaire Small sample size Missing outcome values | Results indicated that a hypothetical intervention intending to increase the distance from children's home to the mines could result in a reduction of rhinoconjunctivitis prevalence in the studied population by up to 4.7 percentage points (95% CI: −8.4%; −1.1%). |
| Pearl M et al. | 2018 | Paediatr Perinat Epidemiol | Life course epidemiology Pregnancy outcomes | Evaluate the separate and combined relations of neighbourhood opportunity in early-life and adulthood with pre-term birth risk in black, white and Latina women + assess the contribution of neighbourhood opportunity to racial-ethnic disparities in risk of PTB. | Longitudinal information | Our results do not point to a common susceptible period across racial/ethnic groups, possibly reflecting unmeasured heterogeneity of experiences represented by the poverty measure, or other unmeasured co-factors. |
| Rudolph KE et al. | 2019 | Environ Epidemiol | Adolescent health | Examine the relation between environmental noise and adolescent health in the US | Missing data - Multiple imputation Sampling weights Propensity score matching Different definitions for exposure Unmeasured confounding / measurement error | We found that living in a community where average day-night noise exceeded the US EPA safety guideline of 55 dB was associated with approximately 30–40 minute later bedtimes. Associations with DSM-IV mental disorders were mixed, generally with wide CIs, and not robust across sensitivity analyses. |
| Casey JA et al. | 2019 | Environ Res | Prenatal exposure | Do lower SES pregnant women have a heightened response to UNGD activity due to coexposure to other environmental and social stressors? | Missing data - Multiple imputation Exposure and outcome measure Assumptions discussed/checked | Our findings revealed an association between living in the highest quartile of a cumulative metric of UNGD activity during pregnancy and increased risk of antenatal anxiety or depression. This increased risk, however, did not appear to mediate the observed association between UNGD activity and preterm birth or reduced term birth weight, as we found no relationship between antenatal anxiety or depression and these outcomes in our sample. |
| Papadopoulou E et al. | 2019 | Environ Health Perspect | Prenatal exposure Biomarkers | Study the association between diet and measured blood and urinary levels of environmental contaminants in mother–child pairs from six European birth cohorts | Variable selection | We estimated that adherence to the dietary recommendations (pregnant women: ≤3 servings=week, children ≤2 servings=week) for fish intake would result in lower exposure to PFASs, As, and Hg compared with those exceeding these recommendation. Fruit consumption was associated with increased levels of urinary OP metabolites concentrations in both pregnant women and children. Using TMLE analysis, we found that consuming more than 2 fruits/d could increase the exposure of pregnant women to OPs, compared with lower fruit intake. |

| Author | Year | Journal | Topic | Objective | Limitations | Findings |
|---|---|---|---|---|---|---|
| Nardone A et al. | 2021 | Environ Health Perspect | Historical epi | Assess the association between HOLC grade and 2010 normalized difference vegetation index (NDVI), a measure of overall greenness. | | We found evidence of an association between worse historical HOLC grade and less 2010 greenspace using data from 102 U.S. urban metropolitan areas. |

**Health Economy**

| Author | Year | Journal | Topic | Objective | Limitations | Findings |
|---|---|---|---|---|---|---|
| Berkowitz S et al. | 2019 | Prev Chronic Dis | Financial resources and food insecurity, healthcare expenditure | Estimate the association between state- and county-level health care expenditures and food insecurity. | Unreliable assumptions in GLMs Ecological measures of exposure and outcome estimated | Adults who were food insecure had annual health care expenditures that were $1,834 (95% CI, $1,073–$2,595) higher than adults who were food secure (P < .001). In children, the model-based estimate for health care costs associated with food insecurity was $80 annually, but this finding was not significant (P = 0.53, 95% CI, –$171 to $329). |

**Non-Communicable Disease Epidemiology**

| Author | Year | Journal | Topic | Objective | Limitations | Findings |
|---|---|---|---|---|---|---|
| Legrand M et al. | 2013 | Crit Care | Post-operative, surgery, kidney | To determine the risk factors for post-operative acute kidney injury in patients operated on for inefective endocarditis. | Multifactorial reasons for the exposure-outcome association | Post-operative AKI following cardiopulmonary bypass for IE results from additive hits to the kidney. We identified several potentially modifiable risk factors such as treatment with vancomycin or aminoglycosides or pre-operative anemia. |
| Leslie HH et al. | 2014 | PLoS One | Hormonal contrception, cervical cancer | To determine whether potential risk factors such as hormonal contraception are true causes for cervical cancer amongst human immunodeficiency virus-positive women in developing countries. | Timing and duration of the exposure, type and measurement of the outcome. | Although selected results suggest an increased prevalence of cervical intraepithelial neoplasia 2 or greater (CIN2+) associated with combined oral contraceptive (COC), evidence is insufficient to conclude causality. |
| Mosconi L et al. | 2018 | PLoS One | Menopause, alzheimer's disease, cognitive performance | Examines the impact of the menopausal transition on Alzheimer's disease biomarker changes and cognitive performance in midlife | Alzheimer's disease is progressive and cross-sectional studies does not capture the ongoing AD process | the optimal window of opportunity for therapeutic intervention to prevent or delay progression of Alzheimer's disease endophenotype in women is early in the endocrine aging process. |

| Author | Year | Journal | Topic | Aim | Limitations | Findings |
|---|---|---|---|---|---|---|
| Torres JM et al. | 2018 | Int J Epidemiol | Adult child migrant, mental health | We examined longitudinal associations between having an adult child migrant and mental health, for middle-aged and older Mexican adults accounting for complex time-varying confounding. | A doubly robust, semiparametric estimation strategy that reduces reliance on correct specification of multiple parametric models. Adjust for time-varying confounders affected by prior exposure | Women with at least one adult child in the US had a higher adjusted baseline prevalence of elevated depressive symptoms (RD: 0.063, 95% CI: 0.035, 0.091) compared to women with no adult children in the US. Men with at least one child in another Mexican city at all three study waves had a lower adjusted prevalence of elevated de- pressive symptoms at 11-year follow-up (RD: 0.042, 95% CI: 0.082, 0.003) compared to those with no internal migrant children over those waves. |
| Yu YH et al. | 2019 | Am J Epidemiol | Obesity, stillbirth | Estimated the association between incident obesity and stillbirth in a cohort constructed from linked birth and death records in Pennsylvania (2003–2013) | Time of onset of obesity is unknown, unclear whether estimated associations are driven largely by persons with long-standing obesity, recently developed obesity, or combination of both. Cause-effect interpretations rely heavily on assumptions if exposure cannot be randomised | we found positive associations of incident prepregnancy obesity with stillbirth |
| Lim S et al. | 2019 | Caries Res | Soda in-take, pediatric caries | Effect of soda intake on dental caries in young children (birth to 5 years). | longitudinal, time-varying data | consistent soda intake during the early childhood led to one additional caries tooth surface. |
| Gianfrancesco MA et al. | 2019 | J Rheumatol | Smoking, rheumatoid arthritis | Examined the association between smoking and rheumatoid arthritis. | Heterogeneous study designs, measurement error in key variables, biases in statistical anaylsis | Smoking is associated with higher levels of disease activity in RA. |
| Mozafar Saadati H et al. | 2020 | Obes Sci Pract | BMI, obesity, stroke, diabetics | Compared the effects of body mass index and central obesity on stroke in diabetics and non-diabetics. | Model misspecification, different distributions of the exposure between covariates | Among diabetics, body shape index and waist-to-hip ratio indices were associated with a higher incidence of stroke. |
| Veit C et al. | 2020 | BMC Med Res Methodol | Asthma, control medication | Calculate the long-term risk of reporting asthma symptoms in relation to control medication use in a real-life setting from childhood to adulthood. | RCT may not represent the general population | we did not observe a beneficial effect of asthma control medication on asthma symptoms. |

36| Author | Year | Journal | Keywords | Objective | Limitations | Conclusions |
|---|---|---|---|---|---|---|
| Yu YH et al. | 2020 | Obstet Gynecol | Newly overweight and obesity, stillbirth | Identify the association of newly developed prepregnancy overweight and obesity with stillbirth and infant mortality. | Time-varying effects of covariates, timing of exposure | Transitioning from normal weight to overweight or obese between pregnancies was associated with an increased risk of stillbirth and neonatal mortality. |
| Decruyenaere A et al. | 2020 | Crit Care | Obesity, survival of critically ill patients | Association between obesity and improved survival among critically ill patients. | Failure to account for confounding and collider stratification bias | TMLE mitigates the obesity paradox observed in critically ill patients, whereas a traditional approach results in even more paradoxical findings |
| Abdollahpour I et al. | 2021 | Am J Epidemiol | Waterpipe smoke (vape), multiple scelrosis | Role of lifetime waterpipe smoking in the etiology of multiple sclerosis (MS). | Estimate marginal effects | these results suggest that waterpipe use, or strongly related but undetermined factors, increases the risk of multiple sclerosis. |
| Reilly ME et al. | 2021 | Foot Ankle Int | Symptomatic hallux valgus, clinical and radiographic outcomes, lapidus procedure vs scarf osteotomy | To compare clinical and radiographic outcomes between patients with symptomatic hallux valgus treated with the modified Lapidus procedure versus scarf osteotomy. | Positivity violations if not appropriately accounted for | Although the modified Lapidus procedure led to a higher probability of achieving a normal intermetatarsal angle, both procedures yielded similar improvements in 1-year patient-reported outcome measures |
| Torres LK et al. | 2021 | Thorax | acute respiratory distress syndrome, mortality, | To estimate the attributable mortality, if any, of acute respiratory distress syndrome (ARDS). | Adequately adjusted for confounders, and utilisation of statistical methodology to estimate causal effects | Acute respiratory distress syndrome has a direct causal link with mortality. |
| Mozafar Saadati H et al. | 2021 | Diabetes Metab Syndr | obesity, cardiovascular disease | To elucidate the effect modification of general and central obesity by sex and age on the risk of cardiovascular events. | Weakness and misspecification of statistcal models. Mediator effect of some biologic factors. | Among males and age 54, waist-to-hip ratio index was associated with a higher risk of coronary heart disease and heart failure while body mass index was so for females and age>54. |
| Almasi-Hashiani A et al. | 2021 | BMC Public Health | reproductive factors, breast cancer | To estimate the causal effect of reproductive factors on breast cancer risk in a case-control study. | Misspecification of regression models may cause ex-treme bias in treatment effect estimates. | Postmenopausal women, and women with a higher age at first marriage, shorter duration of breastfeeding, and history of oral contraceptive use are at the higher risk of breat cancer |

Smith ET AL.

| Author | Year | Journal | Keywords | Objective | Limitations/Considerations | Findings |
|---|---|---|---|---|---|---|
| Dadi AF et al. | 2021 | BMC Pregnancy Childbirth | perinatal depression, infant diarrhea, acute respiratory infection, malnutrition | To estimate associations between perinatal depression and infant diarrhea, Acute Respiratory Infection (ARI), and malnutrition in Gondar Town, Ethiopia. | Differences in findings might potentially be due to bias as a result of unobserved confounding, which cannot be overcome using standard regression techniques. Failure to account for unobserved confounding can weaken the quality of evidence derived from such studies | There was no evidence for an association between perinatal depression and the risk of infant diarrhea, acute respiratory infection, and malnutrition amongst women in Gondar Town. |
| Zou R et al. | 2021 | Clin Nutr | maternal folate levels, pregnancy, neuropsychiatric disorders, | To estimate the association between prenatal exposure to folate and brain development (neuropsychiatric disorders) in late childhood has been rarely investigated. | Appropriately control for confounding | Low maternal folate levels during pregnancy are associated with altered offspring brain development in childhood, suggesting the importance of essential folate concentrations in early pregnancy. |
| Goel AR et al. | 2021 | J Am Acad Audiol | hearing aids, audiometric outcomes | To examine how hearing aids affect standard audiometric outcomes over long-term periods of follow-up. | Reduce bias due to study designs that have found differing evidence | Our analysis revealed discernible effects of 5 years of hearing aid use on hearing ability, specifically as measured by the PTA3-Freq, novel PTAExt, and WRS, suggesting a greater decline in hearing ability in patients using hearing aids |
| Beydoun HA et al. | 2021 | Sci Rep | brain metastases, stereotactic radiosurgery, | To compare hospitalization outcomes among US inpatients with brain metastases who received stereotactic radiosurgery (SRS) and/or non-SRS radiation therapies without neurosurgical intervention | Model misspecification | Stereotactic radiosurgery (SRS) alone or in combination with non-SRS therapies may reduce the risks of prolonged hospitalization and non-routine discharge among hospitalized US patients with brain metastases who underwent radiation therapy without neurosurgical intervention. |
| Chavda MP et al. | 2022 | J Crit Care | obesity, mortality, cardiac arrest | To estimate the conditional and causal effects of obesity on mortality in cardiac arrest patients using the Australian and New Zealand Intensive Care Society (ANZICS) Adult Patient Database (APD). | Methodological issues leading to conflicting results. | After adjustment, there was no association between obesity and outcomes in cardiac arrest patients admitted to intensive care unit. |

| Author | Year | Journal | Topic | Research Question | Sources of Bias | Conclusion |
|---|---|---|---|---|---|---|
| Crowner JR et al. | 2022 | Ann Vasc Surg | chronic limb threatening ischemia, nonoperative management | To assess whether chronic limb threatening ischemia (CLTI) objective performance goals (OPGs) could be attained with nonoperative management alone amongst patients with CLTI. | . | A comprehensive set treatment goals and expected amputation free survival outcomes can guide revascularization, but also assure that appropriate outcomes are achieved for patients treated without revascularization. |
| Akosile M et al. | 2018 | Int J Clin Biostat Biom | survival, right heart catheterisation | Investigated differences in survival among patients with and without right heart catheterisation using data from the Study to Understand Prognoses and Preferences for Outcomes and Risks and Treatments (SUPPORT) | Existence of unadjusted bias in the original analysis | Critically ill patients who received a right heart catheterisation had a significantly decreased 30-day and 60-day survival compared to patients who did not receive one after adjusting for a variety of potential con- founder selection strategies. |
| **Infectious Disease Epidemiology** | | | | | | |
| Schnitzer ME et al. | 2014 | Biometrics | Hepatitis C virus, liver disease | Despite modern effective HIV treatment, hepatitis C virus (HCV) co-infection is associated with a high risk of progression to end-stage liver disease (ESLD) which has emerged as the primary cause of death in this population | identifying and adjusting for variables (baseline or time-varying) that affect both HCV clearance and ESLD

Missing data | We found a clinically but not statistically significant protective effect of the clearance of hepatitis C virus on end-stage liver disease, adjusting for time in the model. |
| Davis FM et al. | 2017 | J Vasc Surg | Surgical site infection, graft failure | Surgical site infection after open lower extremity bypass, leading to increased rate of graft failure | structural or process-of-care character-istics of the hospitals where the procedures were performed | Surgical site infection after lower extremity bypass is associated with an increase in rate of amputation and reoperation |
| Vauchel T et al. | 2019 | Am J Infect Control | Imipenem-resistant acinetobacter baumannii (IR-AB), renal outcomes | To explore the impact of an outbreak of imipenem-resistant Acinetobacter baumannii (IR-AB) on renal outcomes. | . | The episode of imipenem-resistant Acinetobacter baumannii (IR-AB) outbreak was associated with an increased risk of kidney events, which appears to be driven by the use of colistin. |

| Author | Year | Journal | Topic | Background | Rationale | Conclusion |
|---|---|---|---|---|---|---|
| Kempker RR et al. | 2020 | Clin Infect Dis | Drugs for multidrug-resistant tuberculosis | Bedaquiline and delamanid are newly available drugs for treating multidrug-resistant tuberculosis (MDR-TB); however, there are limited data guiding their use and no comparison studies. | limited data on the clinical outcomes of patients treated with bedaquiline and delamanid under programmatic conditions, | Among patients with multidrug-resistant tuberculosis, bedaquiline-based regimens were associated with higher rates of sputum culture conversion, more favorable outcomes, and a lower rate of acquired drug resistance versus delamanid-based regimens |
| Westling T et al. | 2020 | Int J Infect Dis | Sepsis, aqueous chlorhexidine (CHG) used to reduce risk of bloodstream infections (BSI) | Assess the impact of bathing of neonates with 2% chlorhexidine solution on bloodstream infecitons, suspected sepsis, and mortality in a low-income country neonatal care unit. | Causal effects from observational data ncluding time-varying confounders | Aqueous chlorhexidine bathing at admission was associated with a reduced risk of bloodstream infections due to a pathogenic organism after adjusting for potential confounding. |
| Aiemjoy K et al. | 2020 | PLoS Negl Trop Dis | Distance from water source, chlamydia trachomatis, antibody reponses | Children living further from a water source would have higher exposure to Chlamydia Trachomatis and enteric pathogens as determined by antibody responses. | Flexible modelling approach to capture the relationship between seroprevalence and age | Children living further from a water source had higher seropreva- lence of S. enterica and G. intestinalis indicating that improving access to water in the Ethio- pia's Amhara region may reduce exposure to these enteropathogens in young children. |
| Figueroa S et al. | 2020 | Environ Res | Immune stimulation, acute lymphoblastic leukemia | we utilized targeted machine learning and traditional statistical methods to investigate the association of multiple measures of early immune stimulation with acute lymphoblastic leukemia (ALL) in Costa Rican children. | complex biological processes underlying immune dysregulation and leukemogenesis | Exposure to pets and farm animals was inversely associated with acute lymphobalstic leukemia risk, whereas having a fever longer than one week (a putative proxy of severe infection) was associated with an increased risk |
| Amusa L et al. | 2021 | BMC Public Health | male circumcision, sexually transmitted infections | Protective effect of male circumcision against some sexually transmitted infections (STIs) | Mimic an RCT | we present further evidence of a protective association of medical male circumcision against human immunodeficiency virus and herpes simplex virus type-2 in this hyperendemic South African setting. |
| Kerschberger B et al. | 2021 | Am J Epidemiol | anti-retroviral therapy, unfavourable health outcome | Same-day ART, effect on composite unfavourable treatment outcome | Assess real-world effectiveness | There was a reduced risk of composite unfavourable income amongst those with early anti-retroviral therapy initiation comapred to same-day anti-retroviral therapy. |

| | | | | | | |
|---|---|---|---|---|---|---|
| Akhtar S et al. | 2022 | Mult Scler Relat Disord | hepatitis B vaccine, multiple sclerosis | HBV vaccine and multiple sclerosis risk | Marginal causal effect over a population | The results suggest a significant nonspecific protective effect of recombinant Hepatitis B vaccine against multiple sclerosis risk. |
| Chen C et al. | 2022 | J Nutr Health Aging | preoperative hs-crp/Albumin ratio and postoperative SIRS | preoperative hs-CRP/albumin ratio and SIRS | Interactions and stratified analysis | Preoperative hs-CRP/albumin ratio (CAR) was significantly associated with increased risk of postoperative SIRS in elderly patients. |
| Isfordink CJ et al. | 2022 | AIDS | Prevalence hepatitis C virus-viremia, direct-acting antivirals | Prevalence of HCV-viremia. Clinical determinants to lack of direct-acting antivirals | inference to settings with prolonged unrestricted access to treatment | Prevalence of hepatitis C virus-viremic amongst people with HIV is low in the Netherlands, coinciding with widespread DAA-uptake. |
| Figueroa S et al. | 2020 | Environ Res | Immune stimulation, acute lymphoblastic leukemia | we utilized targeted machine learning and traditional statistical methods to investigate the association of multiple measures of early immune stimulation with acute lymphoblastic leukemia (ALL) in Costa Rican children. | complex biological processes underlying immune dysregulation and leukemogenesis | Exposure to pets and farm animals was inversely associated with acute lymphobalstic leukemia risk, whereas having a fever longer than one week (a putative proxy of severe infection) was associated with an increased risk |

**Occupational Epidemiology**

| | | | | | | |
|---|---|---|---|---|---|---|
| Brown DM et al. | 2015 | Epidemiology | Particulate matter, heart disease | incidence of ischemic heart disease (IHD) in relation to accumulated exposure to particulate matter (PM) in a cohort of aluminum workers. | Time-varying confounding, a component of the healthy worker survivor effect | The accumulation of exposure to PM2.5 appears to result in higher risks of ischemic heart disease in both aluminum smelter and fabrication work- ers. |
| Izano MA et al. | 2019 | Environ Epidemiol | Metalworking fluids, colon cancer | the relation between exposure to straight, soluble, and synthetic MWFs and the incidence of colon cancer in a cohort of automobile manufacturing industry workers | Time-varying confounding affected by prior exposure | evidence for a causal effect of straight metalworking fluids exposure on colon cancer risk |

**Pharmacoepidemiology**

| | | | | | | |
|---|---|---|---|---|---|---|
| Sukul D et al. | 2017 | J Invasive Cardiol | pre-procedural P2Y12 inhibitors, percutaneous coronary intervention (PCI). | assess the association between pre-procedural P2Y12 inhibitor administration and clinically important in-hospital outcomes. | . | Decline in the rate of pre-procedural P2Y12 inhibitor administration. No significant differences in outcomes between patients treated with pre-procedural P2Y12 inhibitors and those that were not |

| Author | Year | Journal | Subject | Aim | Limitations | Conclusion |
|---|---|---|---|---|---|---|
| Bell-Gorrod H et al. | 2020 | Am J Epidemiol | human immunodeficiency virus–positive patients, mortality | assessed the impact of delayed switch from first-line ART treatment to second-line ART treatment on mortality in 9 South African treatment programs, a large cohort with long follow-up | Small patient numbers and limited follow-up times in previous studies | Early treatment switch is particularly important for patients with low CD4 counts at failure. |
| Rossides M et al. | 2021 | Respirology | methotrexate or azathioprine. | to estimate the relative risk of infectious disease at 6 months associated with the initiation of methotrexate compared to initiation of azathioprine in patients with sarcoidosis. | | Methotrexate appears to be associated with a lower risk of infection in sarcoidosis than azathioprine, but randomized trials should confirm this finding. |
| Kahkoska AR et al. | 2021 | Diabetes Care | treatment methods after positive test of SARS-COV-2 amongst adults. | determine the respective associations of premorbid glucagon-like peptide-1 receptor agonist (GLP1-RA) and sodium–glucose cotransporter 2 inhibitor (SGLT2i) use, compared with premorbid dipeptidyl peptidase 4 inhibitor (DPP4i) use, with severity of outcomes in the setting of severe acute respiratory syndrome coronavirus 2 (SARS-CoV-2) infection. | Residual confounding | Among SARS-CoV-2–positive adults, premorbid GLP1-RA and SGLT2i use, compared with DPP4i use, was associated with lower odds of mortality and other adverse outcomes, although DPP4i users were older and generally sicker. |
| **Policy** | | | | | | |
| Tran L et al. | 2016 | Epidemiol Methods | Human immunodefficiency virus, low-risk express care task-shifting program | estimating the joint effect of both time to availability of a nurse-based triage system (low risk express care (LREC)) and individual enrollment in the program among HIV patients in East Africa | Point effect of one or more longitudinal exposures, or a series of sequential treatment decisions | small impact of both availability and enrollment in the low risk express care program on incare survival. |
| Skeem JL et al. | 2017 | JAMA Psychiatry | Traditional probation, specialty mental health probation | To test whether specialty probation yields better public safety outcomes than traditional probation. | Short follow-up, limited covariate set | well-implemented specialty probation appears to be effective in reducing general recidivism. |

| Author | Year | Journal | Topic | Objective | Limitations | Conclusions |
|---|---|---|---|---|---|---|
| Skeem JL et al. | 2018 | Psychiatr Serv | Costs or traditional and specialty probation | Specialty mental health probation reduces the likelihood of rearrest for people with mental illness | Small sample size | Well-implemented specialty probation yielded substantial savings—and should be considered in justice reform efforts for people with mental illness. |
| Mehta B et al. | 2021 | Arthroplast Today | Primary care physicians (PCPs) density and total knee and total hip arthroplasty outcomes. | To examine the relationship of primary care provider density with total knee arthroplasty and total hip arthroplasty outcomes. | Missing data, zero inflation | no statistically significant association between PCP density and pain, function, or stiffness outcomes at baseline or 2 years. |
| Wong AK et al. | 2022 | Epidemiology | Public masking mandates, COVID-19 cases and deaths | evaluating the effect of delays in state-level public masking mandates on the relative growth of COVID-19 cases and deaths in the 50 US states from 1 September to 31 October 2020. | Relience on epidemic modelling | public masking mandates are asso-ciated with population-level reductions in COVID-19 spread |
| Moreno-Betancur M et al. | 2022 | Int J Epidemiol | Child development | Evaluate effects on child developmental vulnerability after long-term follow up of a family home visiting program. | Mixed evidence from RCTs, target trial to provide suitable evidence | Results did not provide robust evidence of meaningful beneficial or adverse effects of family home visiting program on child development vulnerability. |
| You Y et al. | 2019 | BMC Med Inform Decis Mak | type 2 diabetes, health programme | assess the performance of a multidisciplinary-team diabetes care program called DIABETIMSS on glycemic control of type 2 diabetes (T2D) patients | Not possible to evaluate a new programme by design, impractical to randomise the initiation | DIABETIMSS program had a small, but significant increase in glycemic control |


    \end{landscape}

\end{document}